\def\grtsim{\mathrel{\hbox{\rlap{\hbox{\lower2pt\hbox{$\sim$}}}\raise2pt\hbox{$>$}}}} 
\def\lesssim{\mathrel{\hbox{\rlap{\hbox{\lower2pt\hbox{$\sim$}}}\raise2pt\hbox{$<$}}}} 
\def\lfir{\hbox{$L_{FIR}$}}
\def\msun{\hbox{$M_{\odot}$}}
\def\lsun{\hbox{$L_{\odot}$}}
\def\kms{\hbox{km s$^{-1}$}}
\def\lco{\hbox{$L'_{CO}$}}
\def\lfir{\hbox{$L_{FIR}$}}
\def\lmir{\hbox{$L_{MIR}$}}
\def\lir{\hbox{$L_{IR}$}}
\def\mh2{\hbox{$M_{H_2}$}}
\def\lsim{\!\!\!\phantom{\le}\smash{\buildrel{}\over
 {\lower2.5dd\hbox{$\buildrel{\lower2dd\hbox{$\displaystyle<$}}\over
                                 \sim$}}}\,\,}
                                 
\documentclass[useAMS,usenatbib]{mn2e}
\usepackage{graphicx}
\usepackage{subfigure}
\usepackage{epsfig}
\usepackage{lscape}
\usepackage{rotating}
\title[Molecular gas in type 2 quasar at $z\sim$0.2-0.3]{Molecular gas in type 2 quasars at $z\sim$0.2-0.3\thanks{Based on observations carried out with the IRAM 30m radiotelescope and the Australia Telescope Compact Array.}}
\author[Villar-Mart\'\i n et al.]{M. Villar-Mart\'\i n$^{1}$, M. Rodr\'\i guez$^2$, G. Drouart$^{3,4}$, B. Emonts$^{5,1}$, L. Colina$^1$ 
\newauthor   A. Humphrey$^6$,  S. Garc\'\i a Burillo$^7$, J. Graci\'a Carpio$^8$, P. Planesas$^7$
\newauthor   M. P\'erez Torres$^2$, S. Arribas$^1$  \\
$^1$Centro de Astrobiolog\'\i a (INTA-CSIC), Carretera de Ajalvir, km 4, 28850 Torrej\'on de Ardoz, Madrid, Spain.  villarmm@cab.inta-csic.es \\
$^2$Instituto de Astrof\'\i sica de Andaluc\'\i a (CSIC), Glorieta de la Astronom\'\i a s/n, 18008 Granada, Spain \\
$^3$European Southern Observatory, Karl Schwarschild Str. 2, D85748, Garching bei M\"unchen, Gemany \\
$^4$Institut dÕAstrophysique de Paris, 98bis boulevard Arago, 75014 Paris, France \\
$^5$CSIRO Astronomy and Space Science, Australia Telescope National Facility, PO Box 76, Epping NSW 1710, Australia \\
$^6$Centro de Astrofisica, Universidade do Porto, Rua das Estrelas, 4150-762 Porto, Portugal \\
$^7$Observatorio Astron\'omico Nacional (OAN)-Observatorio de Madrid, Alfonso XII, 3, 28014, Madrid, Spain \\
$^8$Max-Planck-Institut f\"ur extraterrestrische Physik, Giessenbachstra$\beta$e 1,D-85748,  Garching bei M\"unchen, Germany\\}
\begin{document}

\date{} 

\pagerange{\pageref{firstpage}--\pageref{lastpage}} \pubyear{2002}

\maketitle

\begin{abstract}

We present results of CO(1-0) spectroscopic observations of 10 SDSS type 2 quasars (QSO2) at z$\sim$0.2-0.3 observed with the 30m IRAM radiotelescope and the Australia Telescope Compact Array. We report 5 new confirmed CO(1-0) detections and 1 tentative detection. They have \lco$\sim$ several$\times$10$^9$ K km s$^{-1}$ pc$^2$, while upper limits
   for the  non detections are  \lco $<$ 3 $\sigma$ = several$\times$10$^9$ K km s$^{-1}$ pc$^2$.

 This study increases the total number of QSO2 with CO measurements at $z\la$1 to 20, with a  50\% detection rate.
 The vast majority are at $z\sim$0.1-0.4.
 Assuming a conversion factor $\alpha$=0.8 \msun~(K \kms ~pc$^2$)$^{-1}$, the implied
   molecular gas masses    are in the range    \mh2 $\lesssim$4$\times$10$^8$ to $\sim$5$\times$10$^9$ \msun. 
   We compare with samples of type 1 quasars (QSO1), luminous and ultraluminous infrared galaxies.    We find no difference in the molecular gas content
of QSO1 and QSO2 of a given infrared luminosity, although the QSO2 sample is affected by small number statistics.  This result, if confirmed, is consistent with the unification model for quasars.

 QSO2 fall on the \lco~ vs. $z$,  \lco~ vs. \lfir~ and  $\eta=\frac{L_{ FIR}}{L_{CO}}$ vs. \lfir~correlations defined by  quasars at different $z$.  The location of
   the QSO2 in these diagrams is discussed  in comparison with samples of QSO1, luminous and ultraluminous infrared galaxies, and high $z$ submm sources.

CO(1-0) has FWHM$\sim$180-370 km s$^{-1}$ when detected, with a variety of kinematic profiles (single or double horned). 
In general, the CO line is narrower than [OIII]$\lambda$5007, as observed in low $z$ QSO1, with $FWHM_{ [OIII]}$/$FWHM_{ CO}\sim$1-2). This probably reveals different spatial sizes and/or geometry   of the ionized and molecular phases and a higher sensitivity of the [OIII] emission to non gravitational motions, such as outflows. Considering the $z\sim$0.1-0.4 range, where CO measurements for both QSO1 and QSO2 exist, we find no difference in $FWHM_{ CO}$ between them,
although this result is tentative. In the unification scenario between QSO1 and QSO2,  this suggests that the distribution of CO gas is not related to the obscuring torus.

\end{abstract}

\begin{keywords}
galaxies: quasars: general; galaxies: evolution;  galaxies:interactions.
\end{keywords}

\label{firstpage}

\section{Introduction}

It has been only in the last decade that radio quiet  type 2 (i.e. obscured) quasars, generally known as  ``type 2 quasars", have been discovered in large quantities  at different wavelengths: X-ray (e.g. Szokoly
et al. \citeyear{szo04}), infrared (e.g. Mart\'\i nez-Sansigre et al. \citeyear{sans06},  Stern et al.  \citeyear{ste05}) and optical (Reyes et al.  \citeyear{rey08}, Zakamska et al.  \citeyear{zak03}).   These authors  have identified nearly 1000 type 2 quasars  (QSO2 hereafter, vs. QSO1 or type 1 quasars) at redshift 0.2$\lesssim z\lesssim$0.8 in the Sloan Digital Sky Survey (SDSS, York et al. \citeyear{york00}) based on their optical emission line properties:  narrow H$\beta$ (full width half maximum, FWHM$<$2000 \kms),  high ionization emission lines characteristic of type 2 active galactic nuclei (AGN) and narrow  line  luminosities typical of QSO1 ($L_{[OIII]\lambda 5007})\grtsim$2$\times$10$ ^8$ L$_{\odot}$).

Based on diverse studies it can be said that the host galaxies of QSO2 are  often  ellipticals with frequent  signatures of mergers/interactions (Villar-Mart\'\i n et al. \citeyear{vm12}, Bessiere et al.  \citeyear{bes12}, Villar-Mart\'\i n et al. \citeyear{vm11a}, Greene et al.  \citeyear{green09}). Very intense star formation activity
 is also frequently found  (e.g. Zakamska.  et al.  \citeyear{zak08}, Lacy et al.  \citeyear{lacy07}, Hiner et al. \citeyear{hin09}).
Ionized gas outflows are an ubiquitous phenomenon (Villar-Mart\'\i n et al.  \citeyear{vm11b}, Greene et al. \citeyear{gre11},  Humphrey et al. \citeyear{hum10}).
 The optical continuum is sometimes polarized, revealing the presence of an obscured non-thermal continuum source (Zakamska et al.  \citeyear{zak05}, Vernet et al. \citeyear{ver01}).   

A fundamental piece of information is  still missing: the molecular gas content of this class of objects has  been very scarcely studied and it is not  known whether the host galaxies of QSO2    contain abundant reservoirs of molecular gas.  This gaseous phase  can provide large amounts of fuel   to form stars  and feed the nuclear black hole. This gas is highly sensitive to the different mechanisms at work during galactic evolution (e.g. interactions and mergers). As such, it  retains relic information about the global history of the systems.

CO, the strongest tracer of molecular gas, has been found  in active galaxies (AGNs) at different $z$ of similar  AGN power as QSO2, i.e., QSO1 and powerful FRII (Fanaroff-Riley II) narrow line radio galaxies (see Solomon \& Vanden Bout \citeyear{sol05}, Omont \citeyear{omont07},  Miley \& de Breuck \citeyear{mil08}, for a global review). 
  The inferred H$_2$ masses are in the range $M_{H2} = \alpha$ \lco $\sim$10$^8$-several$\times$10$^{9}$ \msun~ (where \lco~is the CO(1-0) line luminosity and ~assuming  $\alpha$=0.8 \msun~(K \kms ~pc$^2$)$^{-1}$) (Downes \& Solomon \citeyear{dow98})\footnote{Recent  results imply $\alpha$= 0.6 $\pm$ 0.2 (Papadopoulos et al. \citeyear{pap12})} at low $z$ ($z\la$0.1, e.g. Bertram et al. \citeyear{ber07}, Evans et al. \citeyear{ev05}, Oca\~na Flaquer et al. \citeyear{oca10}) and several$\times$10$^{9}$-10$^{11}$ M$_ {\odot}$ at $z\ga$2  (e.g. Emonts et al. \citeyear{emo11a}, de Breuck et al. \citeyear{breu05}, Ivison et al. \citeyear{ivi11}). 
  The presence of CO has also been confirmed  in several distant QSO2 at $z\ga$3, implying masses $\sim$several$\times$10$^{10}$ M$_{\odot}$ (Schumacher et al. \citeyear{schu12}, Polletta et al.  \citeyear{pol11}, Mart\'\i nez -Sansigre et al.  \citeyear{sans09}).

\begin{table*}
\label{tab:log}      % is used to refer this table in the text
\centering                          % used for centering table
\begin{tabular}{lllllllllll}        % centered columns (4 columns)
\hline                % inserts double horizontal lines
 (1) & (2) & (3) & (4) & (5) & (6) & (7) & (8) & (9)  & (10)  \\ 
 Object  & Run & RA & Dec & $z_{\rm SDSS}$ & $D_L$ & Scale & t$_{exp}$  & $\nu_{obs}$ & rms \\       
		&	&	& & &  (Mpc)	 & (kpc/$\arcsec$) &	(hr)	&  (GHz)  &  \\ \hline
SDSS J0831+07 &  A & 08 31 30.3 & +07 05 59.5 & 0.232 & 1147 &  3.7 & 5.2 &  93.6 &  0.4  \\
SDSS J1044+06 & A & 10 44 26.7 & +06 37 53.8 & 0.210 & 1025 & 3.4 &  6.2 & 95.3 & 0.3  \\
SDSS J1106+03   &  A & 11 06 22.0 & +03 57 47.1 & 0.242 & 1204 & 3.8  & 3.9 & 92.8 & 0.4   \\
SDSS J1301-00   & A & 13 01 28.8 & -00 58 04.3 & 0.246 & 1227 & 3.8 & 6.0 & 92.5 &  0.4 \\ 
SDSS J1344+05  &  A & 13 44 18.7 & +05 36 25.6 & 0.276 &  1399 & 4.2 & 6.2 &    90.3 &  0.2 \\ \hline
SDSS J0028-00 &  B & 00 28 52.87 &  -00 14 33.6 & 0.310 &  1601 & 4.5 &    9.2 & 87.1 & 0.2 \\   
SDSS J0103+00 &  B & 01 03 48.58 & +00 39 35.0 & 0.314   & 1625 & 4.6 &    8.0 & 87.7  &  0.3  \\  
SDSS J0236+00 &  B & 02 36 35.06 & +00 51 26.9 & 0.207 & 1009 & 3.4 & 6.5 &  95.5 &   0.2\\ \hline
SDSS 0025-10 & ATCA & 00 25 31.46 &  -10 40 22.2  &     0.303 & 1559 & 4.5 & 17.0 & 88.4 & 1.0 &   \\ 
SDSS 0217-00 & ATCA &  02 17 58.18 &  -00 13 02.7 & 0.344 &  1808 & 4.9 & 7.5 &  85.8 & 2.2 &  \\ 
\hline
\end{tabular}
\caption{The  sample. Objects observed during runs A and B  with the IRAM 30m  radiotelescope 
and with ATCA are separated by  horizontal lines.  $z_{\rm SDSS}$ (5) is the optical redshift derived from the [OIII]$\lambda$5007 line using the SDSS spectra. $D_L$ (6) is the luminosity distance in Mpc. t$_{exp}$ (7)  gives the total exposure time per source including calibrations. $\nu_{obs}$ (8)  is the observed frequency of the CO(1-0) transition and rms (9) is the noise determined from channels with 16 MHz ($\sim$50 \kms) spectral resolution. It is given in mK for the IRAM data and in mJy\,beam$^{-1}$\,chan$^{-1}$ for the ATCA data.}
\end{table*}

   \begin{table*}
\label{tab:log}      % is used to refer this table in the text
\centering                          % used for centering table
\begin{tabular}{llllllllllll}        % centered columns (4 columns)
\hline                % inserts double horizontal lines
 (1) & (2) & (3) & (4) & (5) & (6) & (7)  & (8) & (9)  \\ 
 Object  &   log($\frac{L_{[OIII]}}{L_{\odot}}$) & \lco & $M_{H2}$) & $FWHM_{ CO}$ & $FWHM_{ [OIII]}$ & $V_{CO - [OIII]}$ & \lir & \lfir \\       
		 &  & ($\times$10$^9$) &  ($\times$10$^9$ \msun) &  (km s$^{-1}$) & (km s$^{-1}$) & (km s$^{-1}$) & $\times$10$^{11}$ \lsun  & $\times$10$^{11}$ \lsun \\ \hline
SDSS J0831+07 &  8.52 &  6.5$\pm$1.0 & 5.2 $\pm$0.9  & 370$\pm$60 & 685$\pm$10 &  20$\pm$30 &  $\le$4 & $\le$ 2.6  \\
SDSS J1044+06 &  8.17  &     $<$2.0& $<$1.6  &  - & 1050$\pm$20 &  - &  $\le$2.3  &  $\le$1.6 \\
SDSS J1106+03   &   9.02 &       $<$4.3 &  $<$3.4 & - & 545$\pm$10 &  - &    8.7$\pm$0.6 & 5.7$\pm$0.5    \\
SDSS J1301-00   & 9.14 &     $<$4.1  &  $<$3.3  & - & 760$\pm$10 & - &  $\le$9.5 &  $\le$6.5   \\ 
SDSS J1344+05  &   8.12 &  2.4$\pm$0.6 & 1.9$\pm$0.5   & 180$\pm$30 & 500$\pm$15 &  -310$\pm$50  &  $\le$5.5 & $\le$5  \\ 
   &  &  3.8$\pm$0.6 &  3.0$\pm$0.4 & 220$\pm$30 &    & -20$\pm$30  \\ \hline
SDSS J0028-00 &  8.43  & 6$\pm$1& 5$\pm$1 & 300$\pm$100 &  330$\pm$10  & 60$\pm$30  & - & - \\   
SDSS J0103+00 &  8.31  & 6$\pm$1 & 4.8$\pm$0.9   &280$\pm$50 & 355$\pm$10 &  8$\pm$30  & $\le$18 &  $\le$12 \\  
SDSS J0236+00? &   9.20  & 5.0$\pm$0.5  & 1.7$\pm$0.4 & 220$\pm$40  & 800$\pm$20 & -1670$\pm$30 &  4.2$\pm$1.5 & 2.8$\pm$0.1  \\ \hline
SDSS 0025-10 &  8.73   &   4.3$\pm$0.9  & 3.4$\pm$0.7 & 140$\pm$25  & 440$\pm$10 & -155$\pm$25 & 11.0$\pm$3 $\pm$ & 7.8$\pm$2.2\\ 
   &   & 3.1$\pm$0.9 &  2.5$\pm$0.7 & 80$\pm$40 &    & -5$\pm$20   \\ 
SDSS 0217-00 &  8.81 &  $<$6.2 & $<$5.0 & -&  985$\pm$15 & - & 12.0$\pm$0.8 & 8.7$\pm$0.6  \\
\hline
\end{tabular}
\caption{The luminosity (2) of the [OIII]$\lambda$5007  line (Reyes et al. 2008) is given in log and relative to \lsun.
\lco~   (3) is in units of $\times$10$^9$ K km s$^{-1}$ pc$^2$. A conversion factor $\alpha$=0.8 \msun~(K \kms ~pc$^2$)$^{-1}$ has been assumed to calculate the molecular gas mass (4),  with \msun~= $\alpha\times$ \lco. $V_{CO - [OIII]}$ is the velocity shift of the CO(1-0) line relative to $z_{\rm SDSS}$.
The two values of $FWHM_{CO}$ (5) and  $V_{CO - [OIII]}$  (7)  shown  for SDSS J1344+05 and SDSS J0025-10 correspond to the two spectral components identified in the double horned CO line profile.  SDSS J0236+00 is marked with a ``?" because the detection is only tentative (see \S5.1). The [OIII]$\lambda$5007 FWHM (6) has been measured using the SDSS spectra. Columns (8) and (9) give   the infrared ($\sim$8-1000 $\mu$m) and far infrared (40-500 $\mu$m) luminosities in units of $\times$10$^{11}$ \lsun. \lir~and \lfir ~ could not be constrained for SDSS J0028-00 (see \S4.3).} 
\end{table*}
\begin{table*}
\label{tab:log}      % is used to refer this table in the text
\centering                          % used for centering table
\begin{tabular}{lllll}        % centered columns (4 columns)
\hline                % inserts double horizontal lines
 Object  & $z$ & \lco &  \lir & \lfir   \\ 
   & & ($\times$10$^9$)  & ($\times$10$^{11}$)  & ($\times$10$^{11}$)  \\ \hline
SWIRE2 J021638.21-042250.8  & 0.304  & 2.3 &  2.0$\pm$0.5$^a$ & 1.2$\pm$0.2$^a$ \\
SWIRE2 J021909.60-052512.9   & 0.099 & 0.7 & 0.8$\pm$0.1$^a$ & 0.55$\pm$0.05$^a$ \\
SWIRE2 J021939.08-051133.8  & 0.150 & 1.7 & 0.95$\pm$0.05$^a$ & 0.6$\pm$0.1$^a$ \\ 
SWIRE2 J022306.74-050529.1 & 0.330  & 3.3 & 6$\pm$2$^b$ & 4.0$\pm$1.3$^c$  \\
SWIRE2 J022508.33-053917.7 & 0.293  & $\le$2.2 & 1.5$\pm$0.5$^a$ & 1.2$\pm$0.2$^a$ \\
SDSS J092014.11+453157.3 & 0.403  & $\le$2.8 & 8.2$^d$ & 5.5$^c$ \\
SDSS J103951.49+643004.2?  & 0.402  & 2.1 & 13.0$^d$ & 8.7$^c$ \\
SSTXFLS J171325.1+590531 & 0.126 & $\le$0.5 & 0.23$\pm$0.03$^e$ & 0.15$\pm$0.02$^c$ \\ 
SSTXFLS J171335.1+584756 & 0.133 &  0.5 & 1.0$\pm$0.2$^a$ & 0.7$\pm$0.2$^a$ \\
SSTXFLS J172123.1+601214 & 0.325  &  $\le$1.8 &  4.5$\pm$1.5$^b$ & 3$\pm$1$^c$ \\
\hline
\end{tabular}
\caption{The  KNC12 sample. \lco~ from Table 4 in Krips, Neri \& Cox (2012). \lir~and \lfir~  values: $^a$derived with our SED fitting technique; 
$^b$inferred from equations A22, A23, A24 in Wu et al. (2010). The errors account for the three scenarios described by the equations, but not for the scatter in each formula; $^c$\lfir~obtained from \lfir=\lir/$\xi$, with $\xi$=1.5 (see text); $^d$from Zakamska et al. (2008). No errors quoted by the authors;  $^e$from  Table 2 in Wu et al. (2010).
 A ? indicates a tentative CO(1-0) detection. } 
\end{table*}

Many of  these
  studies have  focussed   at low  ($z<$0.1) and high redshift ($z>$2), and frequently on luminous infrared sources.
The intermediate $z$ range, which spans $\sim$60\% of the age of the Universe, an epoch of declining cosmic star formation rate 
(Hopkins \& Beacom  \citeyear{hop06}) has remained  almost practically unexplored until very recently.  On this regard, two relevant papers have been published recently: on one hand, \cite{xia12}  report CO detections in 17 out of 19  ultraluminous  infrared   QSO1 hosts (10$^{12}$ \lsun $\le$ \lir, where \lir ~is  the infrared luminosity in the $\sim$8-1000 $\mu$m spectral range)   at $z\sim$0.1-0.2. They infer   \mh2$\sim$(0.2-2.1)$\times$10$^{10}$ $\msun$. On the other hand, Krips, Neri \& Cox (2012, \cite{knc12} hereafter) have  investigated for the first time the molecular gas content of 10  QSO2 at $z\sim$0.1-0.4. According to our revised \lir~ values (see \S4.3), all but one have  \lir$\la$several$\times$10$^{11}$ L$_{\odot}$. They confirm the detection of CO(1-0)  in five sources and a tentative detection for a sixth. The derived gas masses are \mh2$\sim$(0.4-2.6)$\times$10$^9$ \msun~for the detections and $\la$2$\times$10$^9$ $\msun$ for the four non detections.\footnote{For coherence with the rest of this work, we have recalculated  \mh2~ for KNC12  sample assuming a conversion factor  $\alpha$=0.8 \msun~(K \kms ~pc$^2$)$^{-1}$ instead of the 4.8 value used by those authors. We have also recalculated their upper limits using the $FWZI$, instead of the $FWHM$ (\S3.3).}   

We present here results on 10 more optically selected QSO2 at $\sim$0.2-0.3 based on data obtained  with the 30m IRAM  radio telescope and the Australian Telescope Compact Array (ATCA). We measure the CO(1-0) luminosities \lco~and constrain the molecular gas masses and the infrared luminosities. These are compared with other samples of quasars,
 luminous infrared galaxies (LIRGs, with 10$^{11}$ \lsun$\le$ \lir $<$ 10$^{12}$ \lsun), ultraluminous infrared galaxies (ULIRGs 10$^{12}$ \lsun $\le$ \lir) and high $z$ submm sources.

 We assume
$\Omega_{\Lambda}$=0.7, $\Omega_M$=0.3, H$_0$=71 km s$^{-1}$ Mpc$^{-1}$.

\section{The sample.}
\label{sec:sam}

The sample consists of 10  radio quiet ($P_{\rm 5GHz} <$10$^{31}$ erg s$^{-1}$ Hz$^{-1}$ sr$^{-1}$,  Miller, Peacock, \& Mead \citeyear{mil90})
 SDSS QSO2 at $z\sim$0.2-0.34 (Table 1) selected from the original sample of \cite{rey08} and  \cite{zak03} (see these papers for a detailed description of the selection criteria). These are   objects with 
narrow (full width half maximum $FWHM\la$1000 \kms) forbidden and permitted emission lines without underlying broad components, with line ratios characteristic of a non-stellar ionizing radiation and [OIII]$\lambda$5007 
 luminosities typical of QSO1.

The IRAM sample consists of 8 SDSS QSO2. No  bias was applied regarding the host galaxy  properties or the infrared (IR) luminosities. The only criteria were  that they were observable with the IRAM radiotelescope and
 with $z\la$0.3, so that the CO(1-0) transition falls within the
 EMIR E090 band. This  transition  is the least dependent on the excitation conditions of the gas, which
 is crucial for deriving reliable estimates of the total molecular gas content, including the wide-spread, low-density gas that may be sub-thermally 
 excited (e.g.  Papadopoulos et al. \citeyear{pap01}, Carilli et al. \citeyear{car10}).

 Two more QSO2 at similar $z$ were observed with ATCA (SDSS J0025-10 and SDSS J0217-00), which were specifically selected for being luminous IR sources (with IRAS counterparts) and for showing interesting features such as being strong mergers. These two systems were  studied in detail by Villar Mart\'\i n et al. (\citeyear{vm11a}, \citeyear{vm11b})  based on deep optical imaging and spectroscopy obtained with the Faint Object FOcal Reducer and low dispersion Spectrograph (FORS2) on the Very Large Telescope (VLT).  A detailed  study of  the CO(1-0) spatial distribution in SDSS J0025-10 based on the ATCA data can be found in  \cite{vm13}.

The luminosity of the [OIII]$\lambda$5007 line, $L_{\rm [OIII]}$,  has been proposed as  a proxy for the AGN power (Heckman et al.  \citeyear{heck04}) and a potential discriminant between  Seyferts and quasars. This is specially useful for type 2 objects for which the optical colours result from a complex mixture of the host galaxy continuum and AGN related sources (e.g. Vernet et al. \citeyear{ver01}).  All but two objects in our sample have  log($\frac{L_{\rm[OIII]}}{L_{\odot}})>$8.3 (Table 2, column 2) which is the lower limit applied by    \cite{rey08}
to select quasars vs. Seyferts.   The two remaining objects have values $\sim$8.1-8.2, i.e., a factor of $\le$1.5 below this limit. However,
taking into account that reddening has been ignored and the fact that [OIII] is partially obscured in type 2 AGNs (di Serego Alighieri et al. \citeyear{dis97}), we consider these two objects as QSO2 as well. It must also be kept in mind that the relation between the [OIII]$\lambda$5007 and bolometric luminosity
for quasars has a significant scatter, resulting in a somewhat arbitrary separation between quasars and Seyferts (e.g. Zakamska et al. \citeyear{zak03}).

We will refer frequently throughout the paper to the QSO2 sample studied by   \cite{knc12}.  All but two objects
 are from the original sample of 24 $\mu$m selected
galaxies observed with the  Spitzer infrared spectrograph for the 5 Millijanksy Unbiased Spitzer Extragalactic Survey (5MUSES) (Wu et al. \citeyear{wu10}, see also Lacy et al. \citeyear{lacy07}).   The other two quasars are from the QSO2 SDSS sample of \cite{zak03}.

\section{Observations}

\subsection{IRAM observations}

The observations were obtained during two different observing runs A and B in February  and August 2012  respectively with the IRAM 30 m telescope at
Pico Veleta, Spain. The EMIR receiver was tuned to the redshifted 
frequencies of the CO line, using the optical SDSS redshift $z_{\rm SDSS}$  for each object (see Table 
1). The observations were performed in the wobbler
switching mode with a throw of 120" (run A) or 50" (run B), in order to 
ensure flat baselines. We observed both polarizations (H and V) using as a 
backend the WILMA autocorrelator that produced an effective total bandwidth 
of 4 GHz with a (Hanning-smoothed) 
velocity resolution of 16 MHz or $\sim$50 \kms.

 \begin{figure*}
\includegraphics{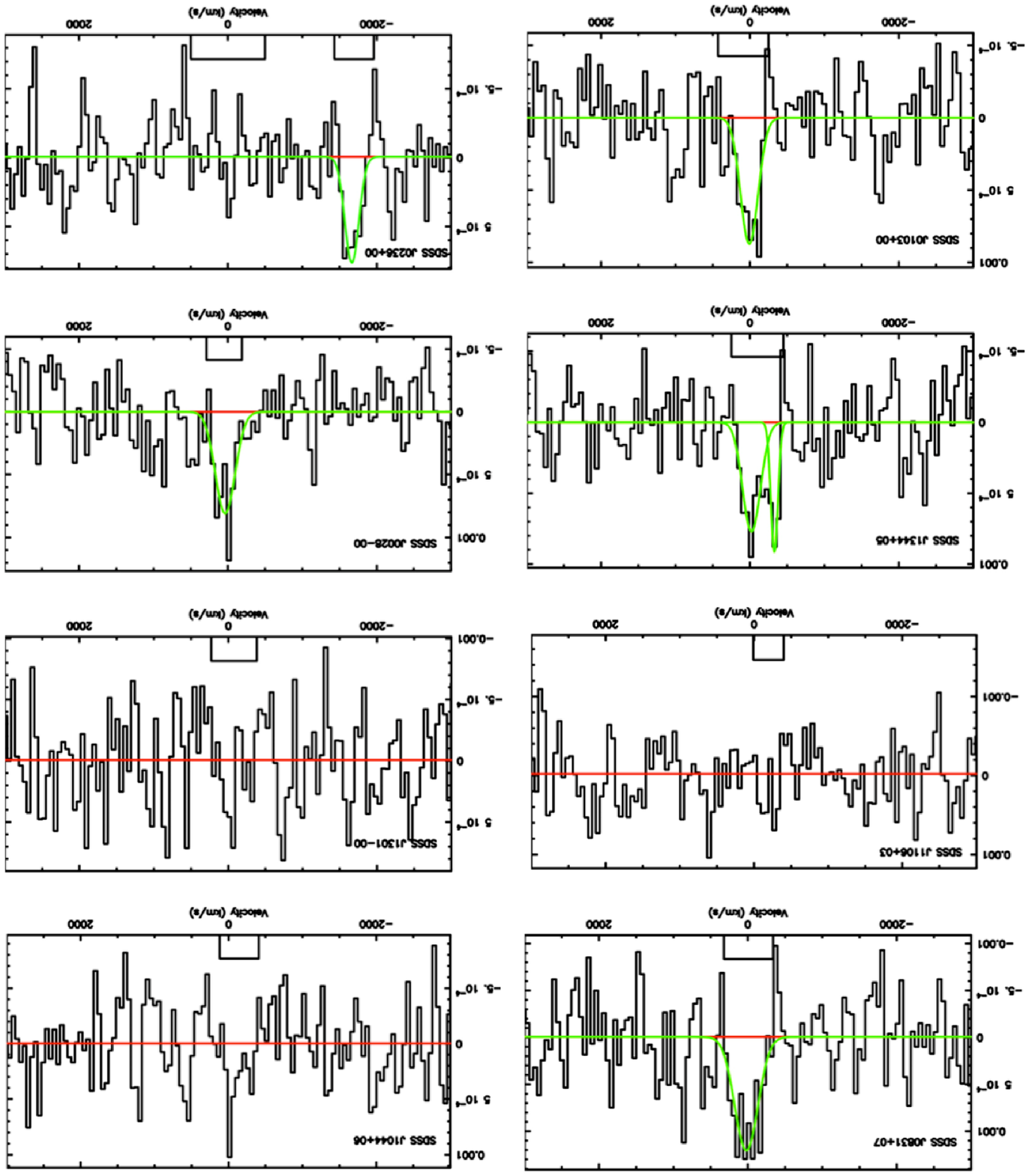}
\vspace{7.5in}
\caption{CO(1-0) spectra of the IRAM sample. The zero velocity corresponds to the optical redshift  $z_{\rm SDSS}$, as derived from [OIII]$\lambda$5007.
Fits of the line profile are shown with green lines for objects with detections or tentative (SDSS J0236+00) detections.  The vertical axis shows  T$^*_A$ in K.}
\end{figure*}

 \begin{figure*}
\includegraphics{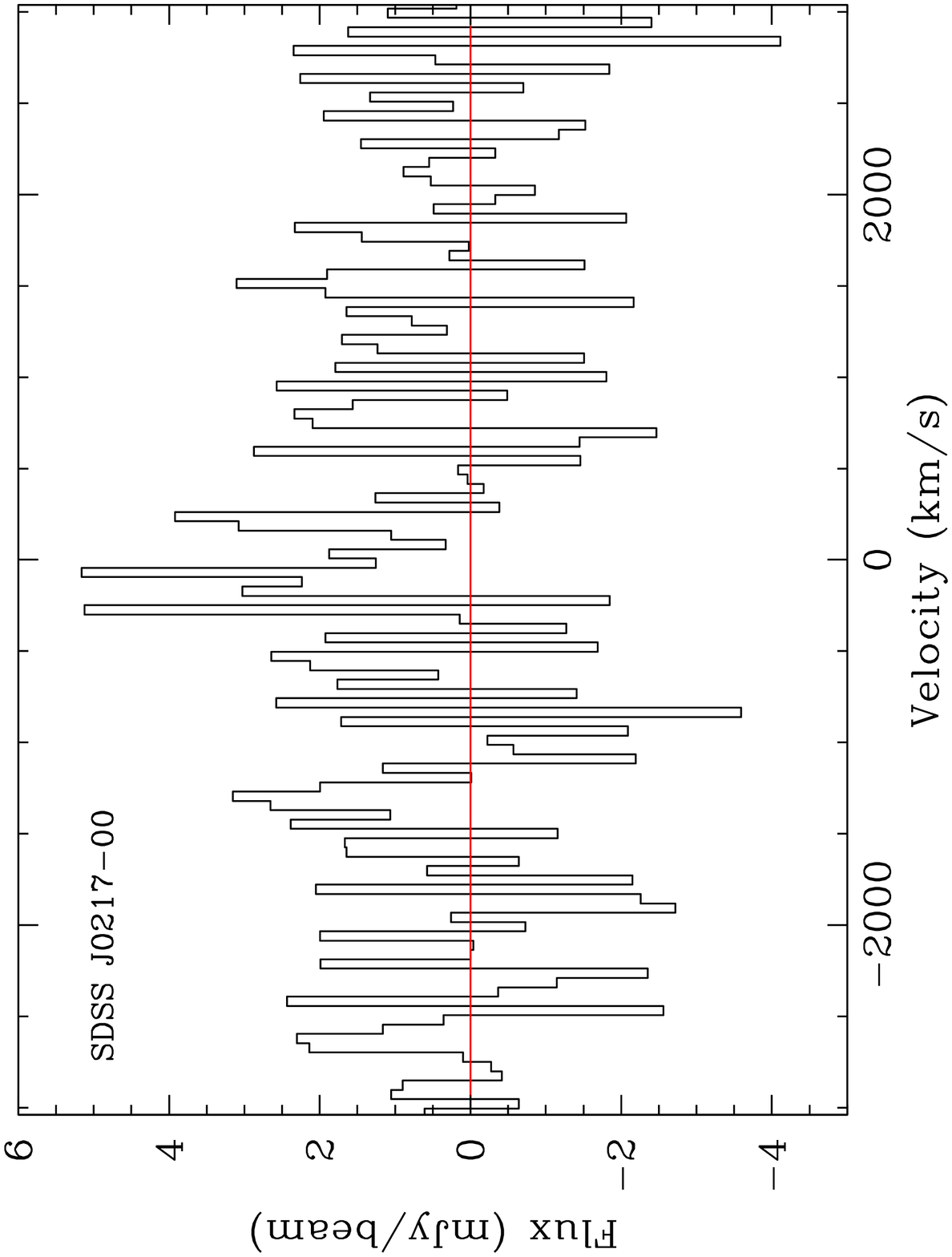}
\includegraphics{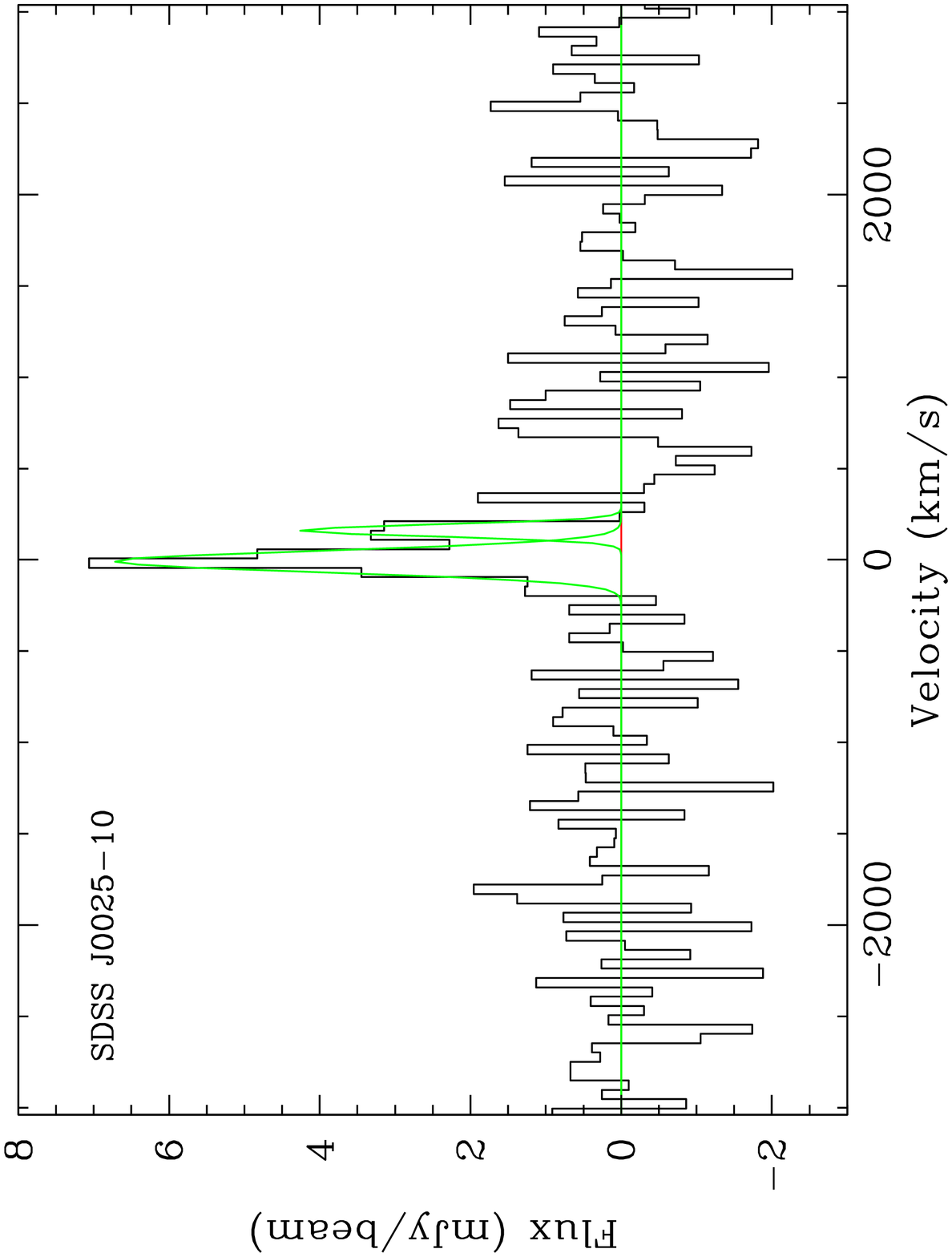}
\vspace{2.5in}
\caption{CO(1-0) spectra of the two QSO2 observed with ATCA. Zero velocity as Fig.~1. Flux units in mJy beam$^{-1}$. Notice the double horned line profile
of SDSS J0025-10.}
\end{figure*}

For run A  the observing conditions were good (pwv $\leq$ 4 mm). For 
run B the conditions were variable with pwv in the range $\sim$3-10 mm. The system temperatures were in the range T$_{sys}\sim$76-114 K for run A
and $\sim$102-106 K for run B. The 
total integration time and the rms for all sources
are specified in Table 1.

The temperature scale used is in main beam temperature T$_{mb}$. At 
3mm the telescope half-power beam width is 29$^{''}$. The main-beam 
efficiency is $\eta_{mb}$= T$^*_{A}$/T$_{mb}$ =0.81. A factor $S$/T$^*_{A}$ = 5.9 Jy/K was applied to obtain the flux in
   mJy/beam  units. 
   
The pointing model was checked against bright, nearby calibrators for every 
source, and every 1.6 hrs for long integrations, it was found to be accurate 
within 5$"$. Calibration scans on the standard two load system were 
taken every 8 minutes. The focus was checked  after sunrise, after sunset and every six hours.

The off-line data reduction was done with the CLASS program of the 
GILDAS software package (Guilloteau \& Forveille \citeyear{gui89}), and 
involved only the subtraction of (flat) baselines from individual
integrations and the averaging of the total spectra.

\subsection{ATCA observations}

The observations of SDSS J0025-10 and SDSS J0217-00 were performed during 2-7 August 2012 with the Australia Telescope Compact Array (ATCA), a radio interferometer in Narrabri, Australia. Observations were done in the compact hybrid H75 array configuration (with baselines ranging from 31 to 89 meters).
Two 2 GHz bands with 1 MHz channel resolution were centered on the redshifted frequency of the CO(1-0) line for each object (88.4\,GHz for SDSS J0025-10  and 85.8\,GHz SDSS J0217-00; Table 1), resulting in a velocity coverage of 7000 km\,s$^{-1}$ and maximum resolution of 3.5 km\,s$^{-1}$. At 88\,GHz, the primary beam of the telescope is 32$^{''}$. Observations were done under good weather conditions, with typical system temperatures of 400 - 650 K and atmospheric seeing fluctuations $< 150$ $\mu$m for SDSS J0025-10 and $< 300$ $\mu$m for SDSS J0217-00 (see Middelberg, Sault \& Kesteven \citeyear{mid06}). The total on-source integration time was 17h for SDSS J0025-10 and 7.5h for SDSS J0217-00.

The phases and bandpass were calibrated every 7.5 minutes with a short ($\sim$2 min) scan on the nearby bright calibrators PKS\,0003-066 (SDSS J0025-10) and PKS\,J0217+0144 (SDSS J0217-00). Atmospheric amplitude variation were calibrated every 30 minutes using a paddle scan, and telescope pointing was updated every hour, or every time the telescope moved $>$20$^{\circ}$ on the sky. For absolute flux calibration, Uranus was observed close to our target source, resulting in an absolute flux calibration accuracy of 20\%.

The off-line data reduction was done with the MIRIAD software. Bad data (including data with internal interference or shadowing of an antenna) were discarded. For the data reduction we followed \cite{emo11a}, noting that atmospheric opacity variations were corrected by weighting the data according to their `above atmosphere' (i.e. paddle-corrected) system temperature. After Fourier transformation, we obtained data cubes with robust weighting +1 \citep{bri95}. 

No 88/85\,GHz radio continuum was detected in these data cubes down to a 5$\sigma$ limit of 0.5 mJy for SDSS J0025-10 and 0.9 mJy for SDSS J0217-00. The synthesized beam-size of the J0025-10 data is $6.30 \times 4.39$ arcsec$^{2}$ (PA -85.3$^{\circ}$) and that of SDSS J0217-00 data is $6.64 \times 5.30$ arcsec$^{2}$ (PA -70.7$^{\circ}$). The spectra presented in this paper were extracted at the location of the quasar host galaxies, binned by 5 channels and subsequently Hanning smoothed to a velocity resolution of 50 km\,s$^{-1}$, yielding a noise level of 1.0 and 2.2 mJy\,beam$^{-1}$\,chan$^{-1}$ for SDSS J0025-10 and SDSS J0217-00 respectively. We also imaged the spatial distribution of the CO(1-0) detected in SDSS J0025-10. This is presented in  \cite{vm13}.

 \section{Analysis}

 \subsection{The comparison samples}

In the following sections (in particular, Figs.~3) we will perform a comparative study between QSO2  and  QSO1, LIRGs,
ULIRGs  and submillimeter sources at different $z$.  

We describe here briefly these samples and the symbol/colour code used in Fig.~3 for their representation. For coherence with our work, upper limits have been recalculated using full width at zero intensity $FWZI$, instead of $FWHM$ whenever possible.

\begin{itemize}

\item QSO1 (blue symbols)

\begin{itemize}

\item non-ULIRG QSO1  at $z\la$0.2 (blue open circles, from Bertram et al. \citeyear{ber07}, Scoville et al. \citeyear{sco03}, Evans et al. \citeyear{ev01}, \citeyear{ev06}, Pott et al. \citeyear{pot06}).

\item  ULIRG QSO1 at $z\sim$0.1-0.3  (blue solid circles, Xia et al. \citeyear{xia12}).

\item  ULIRG QSO1 at $z\sim$0.2-1  (blue solid squares, Combes et al. \citeyear{com12},  \citeyear{com11}).

\item z$\ga$2 QSO1  (blue crosses, Wang et al.  \citeyear{wang11}, \citeyear{wang10}, Coppin et al.  \citeyear{cop08}, Krips et al.  \citeyear{kri05}, Cox et al.  \citeyear{cox02}, Carilli et al.  \citeyear{car02}, Walter et al.  \citeyear{wal04}, Maiolino et al.  \citeyear{mai07}, Gao et al. \citeyear{gao07}, 
Riechers et al.  \citeyear{rie09a},  \citeyear{rie09b}, \citeyear{rie06}).  

\end{itemize}

\item QSO2  (green symbols)

\begin{itemize}

\item 10 QSO2 from this work (green solid circles)

\item 10 QSO2 from KNC12 (green solid squares)

Unlike for QSO1, ULIRG-QSO2  are not included
because as we shall see in \S4.3, the three most luminous QSO2 have \lir~in the LIRG-ULIRG transition regime. All other QSO2 at intermediate $z$ have
\lir~ in the LIRG regime or below.

\item  High $z$ QSO2  (green solid triangles, Polletta et al. \citeyear{pol11}, Mart\'\i nez-Sansigre et al. \citeyear{sans09}, Lacy et al. 
\citeyear{lacy11},  Aravena et al. \citeyear{ara08}, Yan et al. \citeyear{yan10}).

\end{itemize}

\item ULIRGs (no quasars included) and submm sources with no obvious evidence for an AGN (red symbols)

\begin{itemize}

\item  0.04$\la z<$0.2  ULIRGs (red solid diamonds,  Graci\'a Carpio et al.  \citeyear{gra08}).  This sample consists of star forming galaxies  and AGN (Seyfert 1, Seyfert 2 and Liners).

\item 0.2$\la z \la 0.6$ ULIRGs (red solid triangles, Combes et al. \citeyear{com11},\citeyear{com12}).  This sample consists of star forming galaxies  and AGN (Seyfert 1, Seyfert 2 and Liners).

\item $z \ga$2 submm sources with no  obvious evidence of an AGN  (Bothwell et al. \citeyear{bot13}, Ivison et al. \citeyear{ivi11},  Daddi et al. \citeyear{dad09}, Gao et al. \citeyear{gao07}, Weiss et al. \citeyear{wei05}, Smail, Smith \& Ivison \citeyear{sma05},  Neri et al. \citeyear{neri03}, Solomon, Downes \& Radford \citeyear{sol92}).
The luminosities have been corrected for magnification in lensed sources.

\end{itemize}

\item   $z\la$0.05 LIRGs from (orange crosses, Garc\'\i a Burillo et al. \citeyear{gar12},  Graci\'a Carpio et al. \citeyear{gra08}). The LIRG sample consists of  star forming galaxies and AGN (Seyfert 1, Seyfert 2, Liners), but not quasars.

\end{itemize}

\subsection{Calculation of \lco ~and \mh2}

A major part of the results presented in this paper is based on the CO(1-0) line luminosities  \lco ~and the   far infrared luminosity \lfir~  measured between $\sim$40-500 $\mu$m rest frame. To infer them, different methods and observables have been used for different samples and redshifts.  We explain in detail in this and next section  the methodology applied.

 $L'_{CO}$~in K km s$^{-1}$  pc$^2$  is calculated as   (Solomon \& Vanden Bout \citeyear{sol05}):

$$\lco~ = 3.25 ~ \times 10^7 (\frac{S_{CO} \Delta V}{\rm Jy~km/s})~(\frac{D_L}{\rm Mpc})^2 ~(\frac{\nu_{rest}}{\rm GHz})^{-2}(1 + z)^{-1}  $$

where  $I_{CO} = S_ {CO} ~\Delta V$ is  the integrated CO(1-0) line intensity in Jy km s$^{-1}$, $D_L$ is the luminosity distance  in Mpc (Table 1) 
 and  $\nu_{\rm rest}$=115.27 GHz, is the rest frame frequency of the CO(1-0) transition.

   \vspace{0.2cm}

 For the non detections (i.e.  $I_{CO} <3 \sigma$),  we calculate the upper limit as  (Sage \citeyear{sage90}):

$$ I_{CO} < 3~ \sigma_n ~\sqrt{FWZI \times \Delta v} ~~~\rm Jy~km ~s^{-1}  $$

 \vspace{0.2cm} 

where $\Delta v$=50 km s$^{-1}$ is the channel width and   $\sigma_n$ is the channel to channel rms noise of the spectrum in Jy. We have assumed a typical  
$FWZI$ = 870 \kms, using the median value of our $FWZI$ and KNC12 measurements.

Different CO transitions are observable at different $z$. 
In order to extrapolate to the CO(1-0) transition for all objects, we have assumed a constant  effective  brightness temperature for the different transtions (thus, $\frac{\rm L'_{CO(J=n+1\rightarrow n)}}{ L'_{CO(J=n\rightarrow n-1)}}$=1). This is usually assumed for low $z$ studies, where the gas is likely to be optically thick and thermally excited (e.g. Combes et al. 2012; but see also Papadopoulos et al. \citeyear{pap12}). At high $z$ the uncertainties on the CO excitation are larger. Some works suggest that thermal excitation   is a reasonable assumption both for quasars (Riechers et al. \citeyear{rie11}) and  submm sources (Weiss et al. \citeyear{wei05}, Aravena et al. \citeyear{ara08}), while others rather suggest  sub-thermal excitation  (Carilli et al. \citeyear{car10}).  If this were the case,
we would be underestimating the \lco~ by a factor of $\sim$2-4  for the high $z$ sources.  This  and other uncertainties such as  the accuracy of the magnification factor in confirmed lensed objects are likely to contribute to the data scatter at high $z$. However, this will have a small impact on our conclusions since the  scatter of the \lco ~values at a given $z$ is also very large, spanning  $\sim$2 orders of magnitude considering all object classes.

To estimate the molecular gas masses we use the standard conversion formula   (Solomon \& Vanden Bout \citeyear{sol05}):

$$\frac{\mh2}{\msun} = \frac{\alpha}{\msun ~(\rm K ~km ~s^{-1} ~pc^2)^{-1}}~ \frac{\lco}{\rm K~ km ~s^{-1}pc^2}$$

  For the purpose of comparison with other works, we have assumed $\alpha$=0.8 \msun~ (K \kms~ pc$^2$)$^{-1}$, which has been frequently adopted for ULIRGs and active galaxies.

\subsection{Calculation of \lfir}

We plan to investigate the location of our QSO2 in the  \lco~vs. \lfir~ diagram, relatively to other samples. Thus, \lfir~ values are
required for all objets.  Alternatively  \lir~  can be used,
but  \lfir~  is more generally available for the different samples in the literature. Also, it maps cooler dust  
and in principle   it is a more reliable    tracer of the dust emission induced by starburst heating. Finally, \lfir~ is less dependent on orientation than \lir, due to the higher sensitivity of the mid-infrared  luminosity (\lmir = \lir - \lfir) to obscuration (Drouart et al. \citeyear{dro12}).

To constrain  \lfir~ for our QSO2 we have fitted the spectral energy distribution (SED) for the 4 objects with  WISE  (3.3, 4.6, 11.6, 22.1 $\mu$m) and IRAS (60 and/or 100 $\mu$m)  photometric  measurements:  SDSS J1106+03, SDSS J0236+00 and SDSS J0025-10, SDSS J0217-00 (see Figs.~A1 and A2 in the Appendix).  Optical and near infrared photometry have not been used, since these bands are known to be a complex mixture of stellar and AGN related components (e.g. scattered/transmitted AGN light, nebular continuum).  To build the SEDs we used the SWIRE template library (Polletta et al. \citeyear{pol07}) which contains 25 templates including ellipticals,  spirals, starbursts, type 2 and type 1 AGNs and composite starburst + AGN.  The results are shown in Table 2. The uncertainties are dominated by the range of templates able to reproduce the data. SDSS J0028-00  has been excluded in this analysis because an unrelated galaxy very close in projection  confuses the IR photometry.

For the remaining 5 objects in our sample, IRAS upper limits are available and WISE photometry: SDSS J0831+07, SDSS J1044+06, SDSS  J1301-01, SDSS J1344+05, SDSS J0103+00. Only upper limits on \lir~ and  \lfir~  can be obtained for these objects by fitting the SED (Table 2).

Applying the same method, we have recalculated \lfir~ and \lir~ for the 5 objects in KNC12  with both mid and far IR photometric data   to alleviate the large uncertainties  affecting their values (Table 3).  For 4 more objects  \lir~ is available from \cite{wu10} and  \cite{zak08}). For the remaining
 quasar, \lir~was constrained from the 24 $\mu$m luminosity using the equations proposed by \cite{wu10}. To constrain \lfir ~for these quasars, we have estimated a conversion factor $\xi=\frac{\lir}{\lfir}$ appropriate for QSO2. For this, we have used the 9 QSO2 with both \lir~and \lfir{ values available in both samples. They  show very similar $\xi$ in the range 1.4-1.7, with a median value of 1.5, that we assume for $\xi$.   The final revised \lir ~and  \lfir~ values for KNC12 QSO2 are shown  in Table 3.
We have applied the same conversion factor to high $z$ QSO2  with only \lir~ available  (Polletta et al. \citeyear{pol11}) .

To estimate a conversion factor $\xi'=\frac{\lir}{\lfir}$ appropriate for QSO1, 
we have collected the IRAS flux measurements
at 12, 25, 60 and 100 $\mu$m  for  nearby type 1 quasars with the 4 measurements published ($\sim$17 objects in Sanders et al. \citeyear{san89a}). 
 We find that  $\xi'$  is in the range  $\sim$2.0-3.3, with a median value of 2.96, which we therefore assume to estimate \lfir~  for those few QSO1 with no available \lfir.

The low $\xi$ values we have measured for $z<$0.4 QSO2 are consistent with the unification scenario of QSO1 and QSO2.
They suggest that the MIR emission, which is expected to be emitted by the hottest dust in the inner faces of the obscuring torus (Drouart et al. \citeyear{dro12}) is partially obscured. This is consistent with \cite{hin09}, who found that QSO1 have less  far-IR emission on average when compared to QSO2 matched in mid-infrared luminosity. Similarly, the authors propose that this difference is due to orientation.

 We have used a  conversion factor $\xi_* = \frac{\rm L_{IR}}{\rm L_{FIR}}$=1.3 appropriate for  high $z$ submm (Ivison et al. \citeyear{ivi11}) sources with no evidence for an AGN. This is the median value inferred
 for the sample of  $z\la$0.2 ULIRGs of  \cite{gra08}.  It is consistent with   works which show that  most 
 non active galaxies  have $\xi_*\sim$1.3 (Pott et al. \citeyear{pot06}) over  several orders of magnitude of \lir .    \section{Results and discussion}

\subsection{\lco~ and \mh2}

 CO(1-0) detection (S/N$\geq$3 over the integrated line profile) is confirmed in  5 out of the 10 quasars observed (Figs.~1, 2 and Table 2): SDSS J0831+07, SDSS J1344+05, SDSS J0028-00, SDSS J0103+00
 and SDSS J0025-10. For a 6th object,  SDSS J0236+00, we claim a tentative detection. The spectrum shows  an emission line feature detected at 5$\sigma$ level with $FWHM$=220$\pm$40 km s$^{-1}$. The shift in velocity relative to [OIII]$\lambda$5007 is very large, with $V_{CO - [OIII]}$=-1670$\pm$30 km s$^{-1}$ 
 compared with $<$100 \kms~ measured for the other objects (see column (12) in Table 1).  A similar case was discussed by KNC12, although these authors can confirm that the CO emission is close to the  spatial position of the  QSO2 radio emission. 
 
The K-band image of SDSS J0236+00 shows a  disturbed morphology (Stanford et al. \citeyear{sta00}). There is an
 object located at $\sim$5$\arcsec$ ($\sim$17 kpc)  NW of the quasar (thus, well within the 29\arcsec telescope beam) with a hint of a tidal tail connecting it to the quasar. However,
  the $z$ is unknown and it could be an unrelated source. In any case, even if  this object is confirmed to be a  companion, it seems unlikely that the CO line is associated with it, since the velocity shift is rather extreme for a galaxy pair (Patton et al. \citeyear{pat00}).
If the CO emission line feature is not associated with the quasar, we estimate an upper limit for \lco $<$ 3$\sigma$ = 3$\times$10$^9$
K km s$^{-1}$  pc$^2$. 
 
   All 6 quasars with confirmed or tentative CO detection have \lco$\sim$ several$\times$10$^9$ K km s$^{-1}$pc$^2$ (Table 2), while upper limits
   for the 4 non detections are  \lco $<$ 3 $\sigma$ = several$\times$10$^9$ K km s$^{-1}$ pc$^2$. The broad band width ($\pm$5000 km s$^{-1}$ relative to the optical redshift) of the spectra ensures that the non detections are real, rather than due to a shift in velocity of the molecular gas emission out of the observed spectral band. 
    For comparison, the  objects with  definite CO  detections in KNC12   sample   ($z\sim$0.1-0.4) have \lco~in the range (0.5-3.3)$\times$10$^9$ K km s$^{-1}$ pc$^2$, while the non detections have in all cases \lco $<$ 3 $\sigma\sim$3 $\times$10$^9$ K km s$^{-1}$ pc$^2$.

  The implied molecular gas masses for our sample assuming  $\alpha$=0.8 \msun~ (K \kms~ pc$^2$)$^{-1}$ are  $\sim$(2-6)$\times$10$^9$ \msun~ for the quasars with detections and $\la$several$\times$10$^9$ \msun~ for the non detections. Masses $\la$2$\times$10$^9$ \msun~  are derived for all but one QSO2  in KNC12. Their objects have in general lower IR luminosities  which can also explain the lower \mh2 (see \S4.2). For comparison,
  the Milky Way contains $\sim$(2-3)$\times$10$^9$ \msun~of  molecular gas (Combes \citeyear{com91}).

 We conclude that the 20 QSO2 observed so far at $z\sim$0.1-0.4 (KNC12 and our  sample) have CO(1-0) luminosities in the range   \lco~$\lesssim$5$\times$10$^8$ - 6.5$\times$10$^9$ \lsun~  and \mh2 $\lesssim$4$\times$10$^8$ - 5$\times$10$^9$ \msun. Most of these QSO2 have total IR luminosities $<$10$^{12}$ \lsun. Larger molecular gas reservoirs $>$10$^{10}$ \msun~ will probably be found when $\lir >$several$\times$10$^{12}$ \lsun~ QSO2 are investigated. We next compare with other QSO samples, as well as LIRGs and ULIRGs.

\subsection{\lco~vs. $z$, \lco~vs. \lfir~  and $\frac{L_{FIR}}{L_{CO}}$ vs. \lfir}

$L'_{CO}$  is known to correlate both with $z$ and \lfir~  for different types of galaxies, active and non active (e.g. Solomon \& Vanden Bout \citeyear{sol05}).  
These apparent correlations reflect partially a selection bias  since at the highest $z$ we are sensitive only to the most luminous  CO and IR emitters.  
However, this is unlikely the whole story and a combination of the steep decline at the highest luminosities of the \lco~ and \lfir~  luminosity functions and the  evolution of such functions  with $z$ are also
likely to play a role  (P\'erez Gonz\'alez et al. \citeyear{per05}, Keres, Yun \& Young  \citeyear{ker03}, Lagos et al. \citeyear{lag11}).  On the other hand, some works suggest that distant star forming galaxies were indeed much more molecular-gas rich   (e.g. Tacconi et al. \citeyear{tac10}, Daddi et al. \citeyear{dad10}).

As in other galaxy types,  the \lco~ vs. \lfir~  correlation also reflects that more intense star formation is associated with  larger contents of molecular gas.  The interpretation is not  so clear cut in quasars given the uncertain contribution of the AGN to the dust heating. Although \lfir~  is  less affected than \lir~  by this effect due to  the dominant contribution of the AGN to  \lmir~  (see \S4.3), some  contamination  cannot be totally ruled out (e.g. Hiner et al. \citeyear{hin09}). 
 
We show \lco~vs. $z$  and \lco~vs. \lfir~in Fig.~3\footnote{Due to the  uncertainty regarding  the CO detection in SDSS J0236+00, we will use  the \lco~ upper limit in the discussion that follows.}. Quasars only  (blue and green symbols) are included on the left panels (A and B). LIRGs (orange symbols) and ULIRGs and high $z$ submm sources  with no evidence for an AGN (both represented with red symbols) are added on the right panels (C and D).   
The well known \lco~vs. $z$~ and \lco~ vs. \lfir~ correlations for quasars is clearly appreciated in panels (A) and (B) respectively. 

Quasar activity is clearly triggered in systems spanning a range of more than 4 orders of magnitude both in infrared and CO luminosities (or molecular gas content, assuming the same $\alpha$). Fig.~3-C and D show that at low $z<0.1$, all quasars with CO measurements are QSO1.   In general they have lower \lco~ and (for  objects with  \lfir~ available) also 
 lower \lfir~ than LIRGs at similar $z$ and ULIRGs in general.  The scarcity of low $z$ QSO1 with high \lfir~luminosities in the LIRG regime or higher suggest that they are  are intrinsically different from more distant quasars ($z>$0.1) in the diagrams. It is not clear what the difference is. Maybe the increasing incidence of major  vs. minor mergers as  $z$ and/or \lfir~increases. 

Let us focus on the $z\sim$0.1-0.4  range covered by the QSO2 sample.  We have enlarged the total  sample of quasars studied at this intermediate $z$ by KNC12 with  19 ULIRG QSO1 from \cite{xia12} (blue solid circles in Fig.2) and our 10 QSO2 (green solid circles). 

We find that the 20  QSO2  observed so far at intermediate $z$ fall on the \lco~ vs. $z$  and \lco~vs. \lfir~ correlations.   KNC12 found a trend for QSO2 to have lower \lco~
 values than QSO1 at similar $z\sim$0.1-0.4 (Fig.3-A).  
  Adding our sample and the ULIRG QSO1 at similar $z$ demonstrates that this difference  is a consequence of their lower infrared luminosities.  The \lco~median values for our  and KNC12 samples are \lco$^{\rm med}\sim$6.0$\times$10$^9$  and $\sim$2.3$\times$10$^9$ K \kms~ pc$^2$ respectively, taking into account the upper limits. On the other hand,  \lfir$^{\rm med}\sim$10$^{11}$ \lsun~ for KNC12.   \lfir$^{\rm med}$ is rather uncertain for our sample given the numerous upper limits. However,   considering different realistic scenarios about the possible range of values (Wu et al. \citeyear{wu10}),  \lfir$^{\rm med}>$10$^{11}$ \lsun~is always found. Thus,  our sample contains more luminous infrared sources, which  explains the higher  \lco$^{\rm med}$. This is on the other hand somewhat surprising, since  most objects in KNC12  are 24 $\mu$m selected sources,   while our sources were selected in the optical from the SDSS QSO2 database. Some unknown bias (e.g. maybe warmer or less obscured sources in  KNC12) is possibly at work.   
  
 The influence of \lfir~is also clear  in Table 4 where we compare \lco$^{\rm med}$~   and \lfir$^{\rm med}$~  for different samples  of QSO1, QSO2, LIRGs and ULIRGs at $z\la$0.4. It can be seen that samples with similar \lfir$^{\rm med}$ have also similar \lco.

 Thus, for a fixed \lfir, QSO1 and QSO2 at $z\sim$0.1-0.4 are indistinguishable regarding their molecular gas content,  assuming the same conversion factor $\alpha$ applies. 
 Comparison with (U)LIRGs reinforces that differences in \lco~ are a consequence of variations in \lfir.       Although the QSO2 sample is affected by small number statistics, this result, if confirmed, is consistent with the unification model of QSO1 and QSO2.
 
To perform a more complete and adequate comparison  between QSO1 and QSO2 it is  essential  to
  expand this study in $z$ and \lfir, for both QSO1 and very specially   QSO2. The low $z$ ($z<$0.1) range of QSO2 is completely unexplored, as well as the ULIRG regime at intermediate $z$.  Similarly, it will be useful  to enlarge the sample of non-ULIRG QSO1 at $z>$0.2 to ensure an overlap on both $z$ and \lfir~with the QSO2 samples. 

 \begin{table*}
\label{tab:log}      % is used to refer this table in the text
\centering                          % used for centering table
\begin{tabular}{lllllll}        % centered columns (4 columns)
\hline                % inserts double horizontal lines
 Object  & Nr. & $z$ range &   $z^{\rm med}$&  \lfir$^{\rm med}$&  \lco$^{\rm med}$  \\     
class & & & & $\times$10$^{11}$  & $\times$10$^{9}$  & \\  \hline
QSO2 & 20 & 0.1-0.4 & 0.28 & 2.0 &   2.0  \\
non-ULIRG QSO1  & 8 &  0.1-0.2   &  0.15 &   1.8 & 1.2  \\
ULIRGs QSO1  & 19 &  0.1-0.3 & 0.15  & 11.9  & 9.5  \\ \hline
LIRGs &  50 & 0.003-0.05 & 0.017 &  2.1 &  3.0 \\  
ULIRGs & 103 & 0.1-0.4 & 0.17 &   19.4 & 8.9\\ 
\hline
\end{tabular}
\caption{Comparison of \lfir~and \lco~ between QSO2, QSO1, LIRGs and ULIRGs at $z\la$0.1-0.4 (data for LIRGs at $z>$0.05 are very scarce). Nr. is the number of objects in each sample. It is found that samples with similar \lfir$^{\rm med}$~
have similar \lco$^{\rm med}$. For a given \lfir, no distinction is found between QSO1 and QSO2.}
\end{table*}

We show in Fig.~4 (top) \lco~vs. \lfir~  only for quasars (blue squares for QSO1  and green circles for QSO2), excluding upper limits  for coherence with other works.  A non-linear relation \lco $\propto$ \lfir$^{0.68\pm0.10}$ (black solid line) is found  for the QSO1 and QSO2 combined sample (black solid line). The correlation coefficient $R^2$ is 0.89. The fit is consistent with
 the best fit  power law slope found by other authors (Xia et al.  \citeyear{xia12}, \cite{knc12}). When differentiating  QSO2 and QSO1, the slopes are 0.66 and 0.70 respectively, well within the scatter, so it cannot be said whether a real difference is present.  At the highest \lfir$>$10$^{12}$\lsun~ QSO2 seem to lie above QSO1. This difference is also apparent when including the upper limits
 for both object classes (Fig.~3-B).  However, most of  these  objects are at  high $z$, and the measured values are affected by large uncertainties in general and poor statistics for QSO2.  Thus, it is not clear that the difference  is real. This further reinforces the interest of  exploring  the highest \lfir ~ regime for QSO2
  at different $z$.

 As already found by \cite{xia12},  no dependence  is found for the  \lco~vs. \lfir~ slope with \lfir.  \cite{gao04} found that the slope for samples of low luminosity star forming galaxies, LIRGs and ULIRGs becomes steeper as the infrared luminosity increases (the scatter in our samples of ULIRGs and LIRGs is very large and the 
 change of slope might be masked).  This change of slope  has been widely discussed in the literature. 
 Different works propose that this is due to a variation of the relative ratio between the densest molecular gas, responsible for forming stars (traced at least in low \lfir~ systems by HCN)
 and the less dense CO emitting gas (Garc\'\i a Burillo et al. \citeyear{gar12}, Gao \& Solomon \citeyear{gao04}).

 We have computed   $\eta = \frac{L_{FIR}}{L'_{CO}}$.  It is used as a tracer of the star formation efficiency ($SFE = \frac{L_{FIR}}{M_{\rm H2}}$) in objects where the IR luminosity is dominated by starbursts. If there is a significant contribution of the AGN to \lfir~in quasars, then $\eta$ gives an upper limit on the $SFE$.    On the other hand, different works show that using the HCN luminosity instead of \lco~ results in more  reliable $SFE$ values at least for non-ULIRG systems  (Garc\'\i a Burillo et al. \citeyear{gar12}, Graci\'a Carpio et al. \citeyear{gra06}, Gao \& Solomon \citeyear{gao04}).  Therefore, the interpretation of $\eta$ when comparing different samples is not trivial. However, provided these caveats are taken into account, the exercise  can provide useful information,  at least to constrain the exact role of such caveats.

 $\eta$ vs. \lfir~ is shown for quasars in Fig.~4 (middle).    In this case, the vertical axis does not depend on the distance. LIRGs (orange crosses) and also ULIRGs and  high $z$ submm sources (all represented with red diamonds) are added in Fig.~4 (bottom). Objects with upper limits for \lfir~and \lco~have been eliminated as 
 above. Clearly $\eta$ correlates with \lfir~ (as widely discussed in the literature; e.g. Gao \& Solomon \citeyear{gao04}, Xia et al. \citeyear{xia12}). Although the number of
 QSO2 is small, they also follow this trend (middle panel).   At  high \lfir$>$10$^{12}$\lsun, QSO2 apparently tend to lie below QSO1, but see warning above. 

The trend defined by QSO overlaps with that defined by LIRGs and ULIRGs. At a given \lfir, 
there is no clear shift of the QSO  towards higher $\eta$ values relative to other samples of similar \lfir.
Such a shift could be expected  if the \lfir~is contaminated by dust emission heated by the AGN. However, 
the scatter is so large that the increase in $\eta$ might be masked. Alternatively, such contamination might be negligible and thus, 
QSO  would have similar star formation efficiencies than LIRGs or ULIRG of similar infrared luminosities.

  \begin{figure*}
\includegraphics{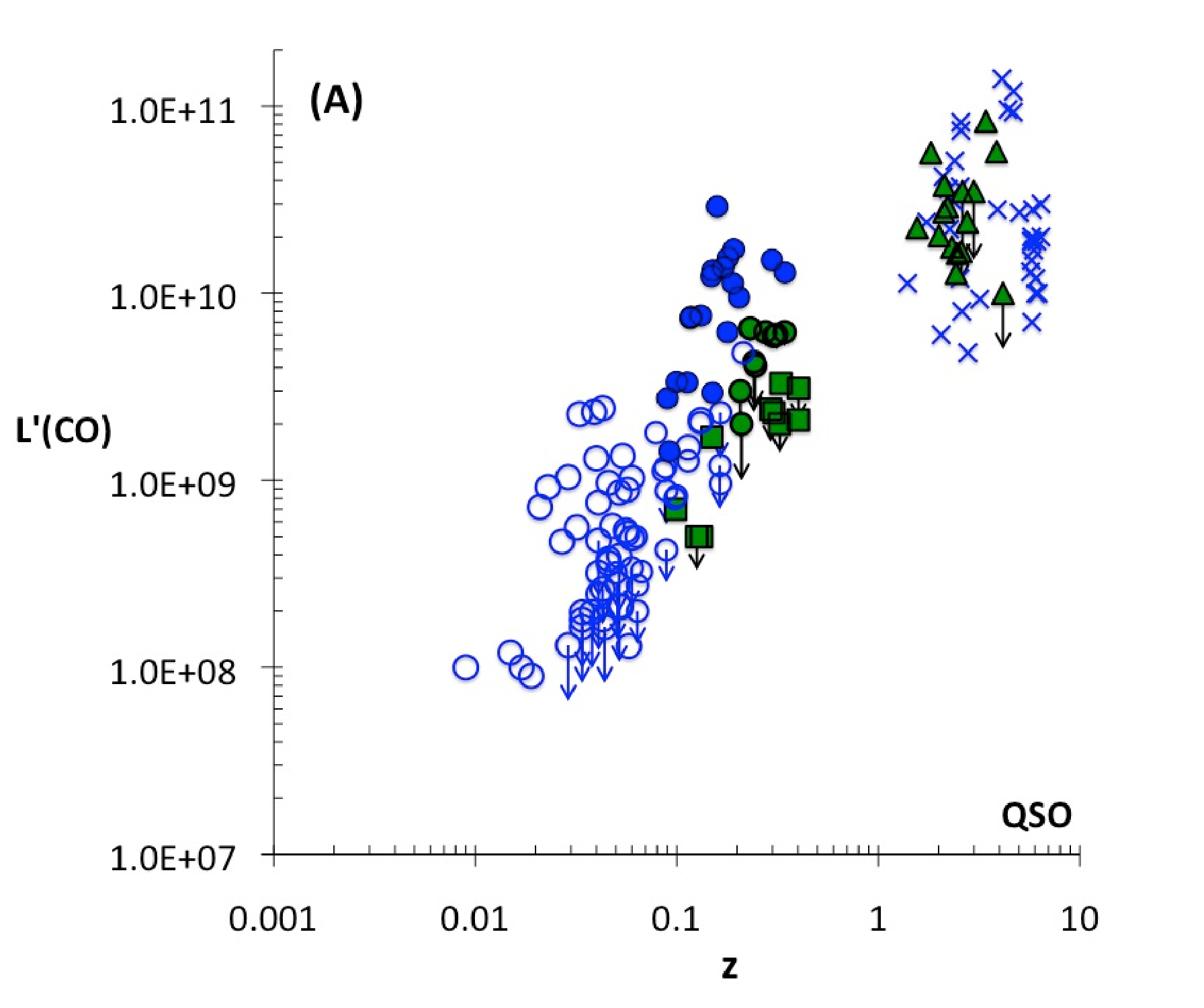}
\vspace{3.7in}
\includegraphics{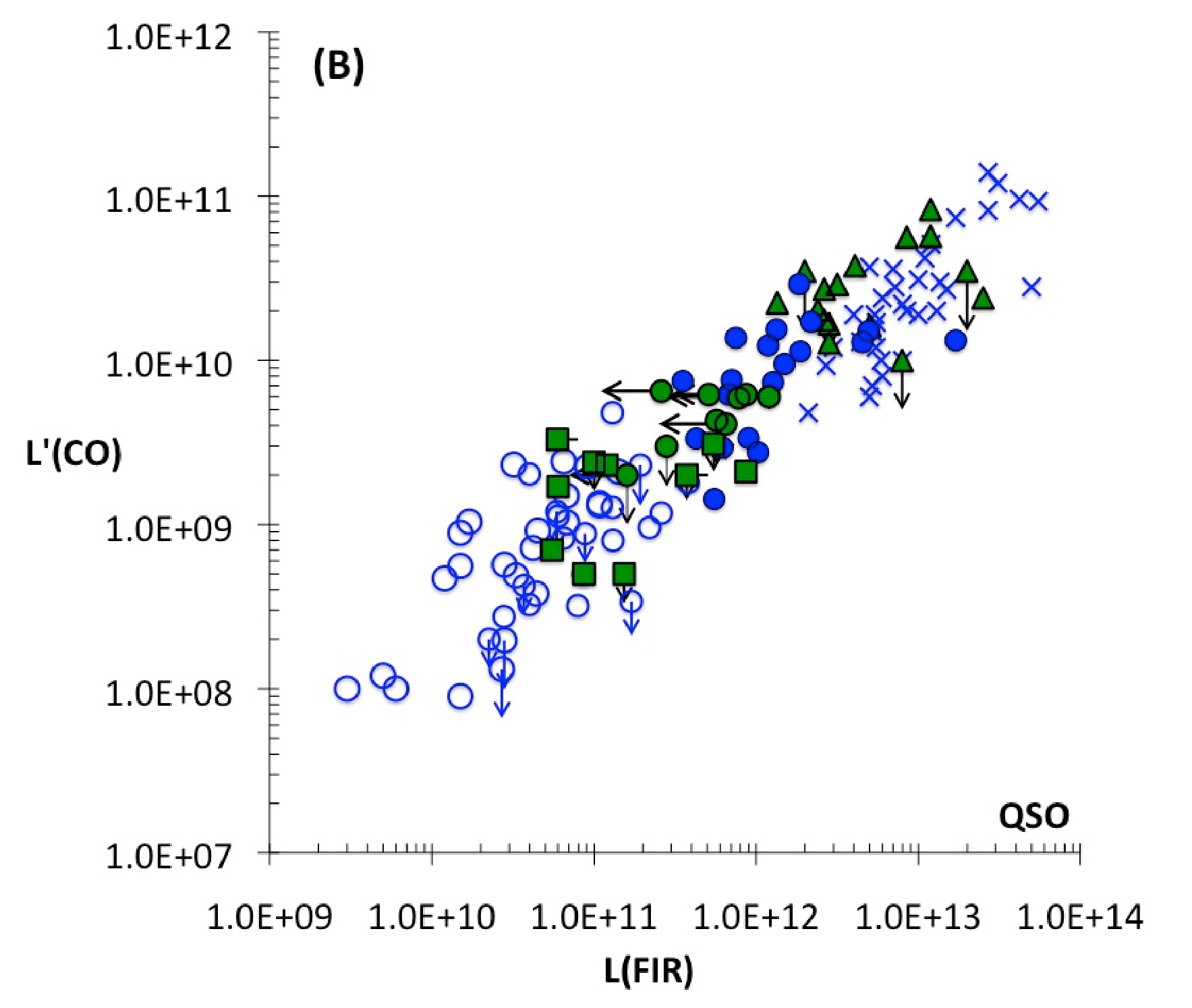}
\vspace{3.5in}
\caption{ \lco~ vs. $z$ (left) and   \lco~ vs. \lfir~(right). The left panels show only quasars: type 1 (blue symbols) and type 2 (green symbols).  LIRGs  (orange crosses) and ULIRGs (red symbols) (no quasars included in these samples) and high $z$ submm sources with no evidence for an AGN are added on the right panels. References: DW (different works);  X12 (Xia et al. 2012); C11 (Combes et al. 2011); C12 (Combes et al. 2012); KNC12 (Krips, Neri \& Cox 2012);  GB (Graci\'a Carpio et al. 2008); GC12 (Garc\'\i a Burillo et al. 2012).}
\includegraphics{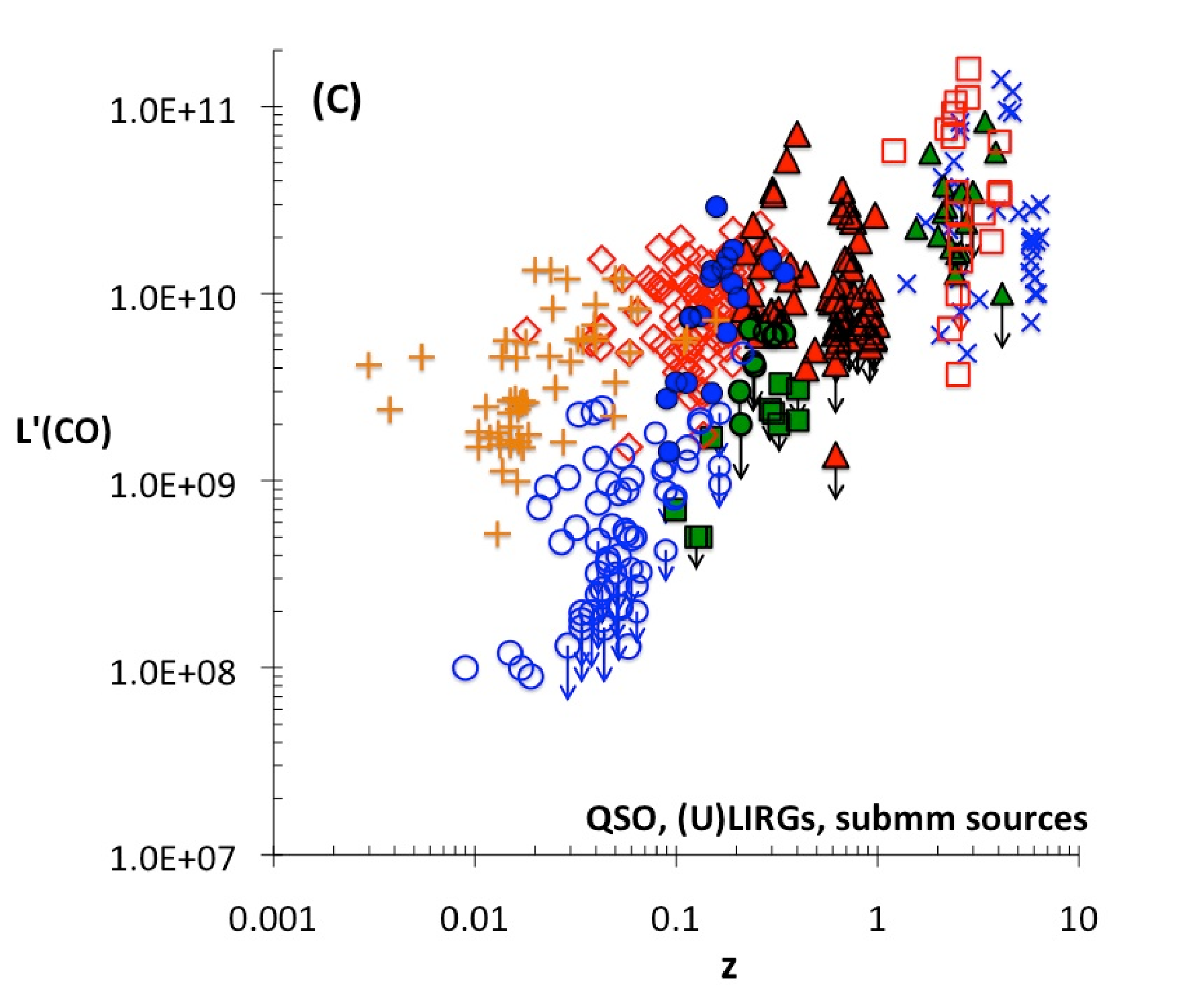}
\vspace{3.7in}
\includegraphics{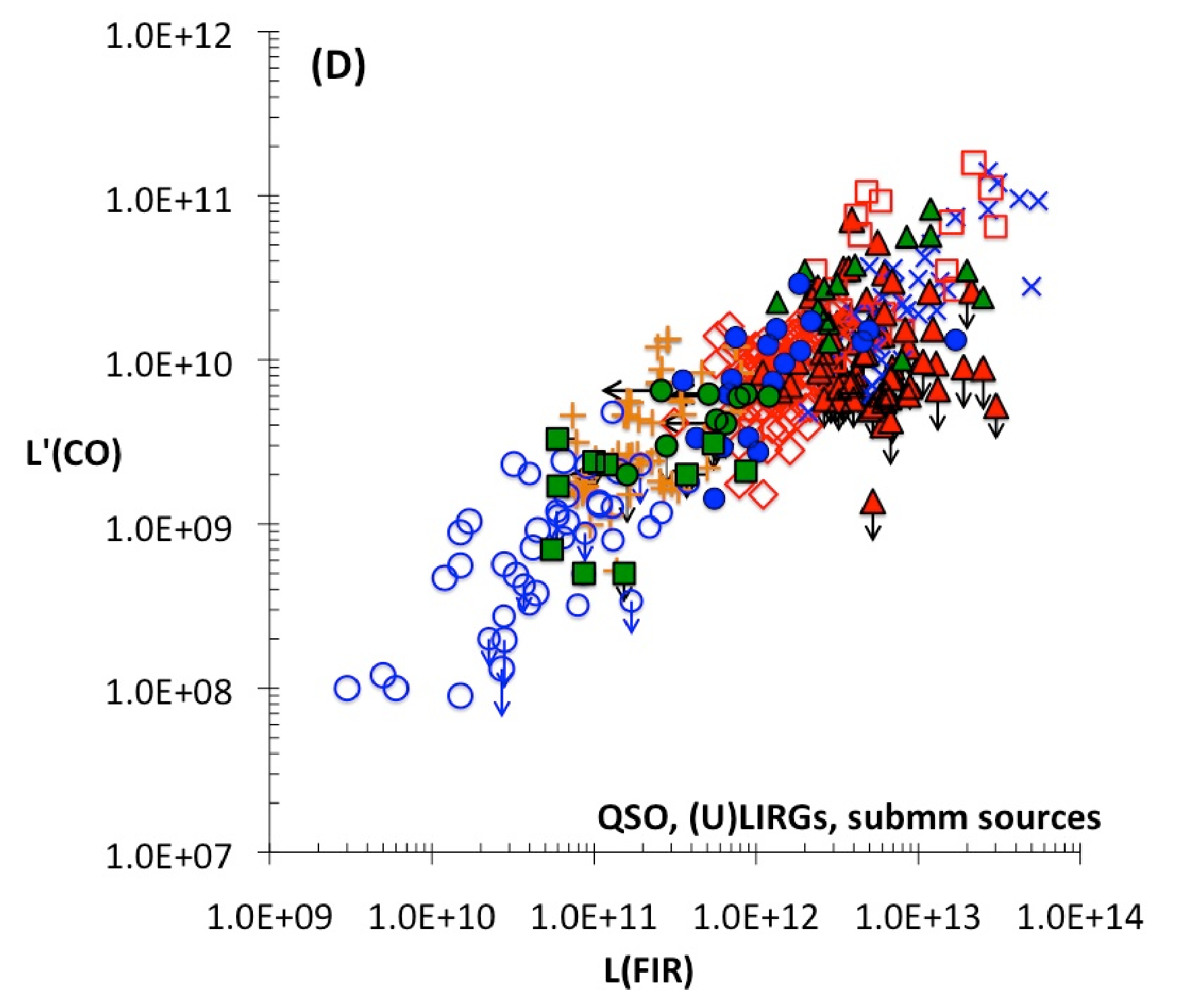}
\vspace{3.5in}
\includegraphics{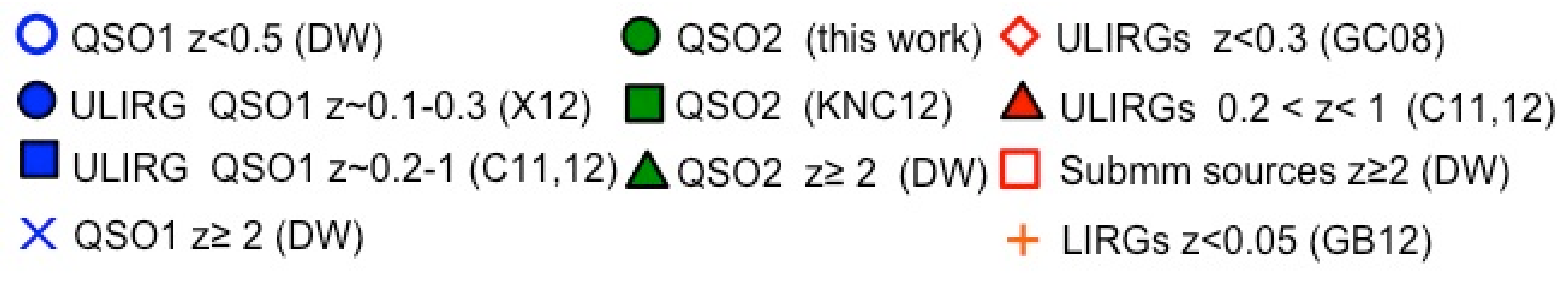}
\end{figure*}

 \begin{figure}
\includegraphics{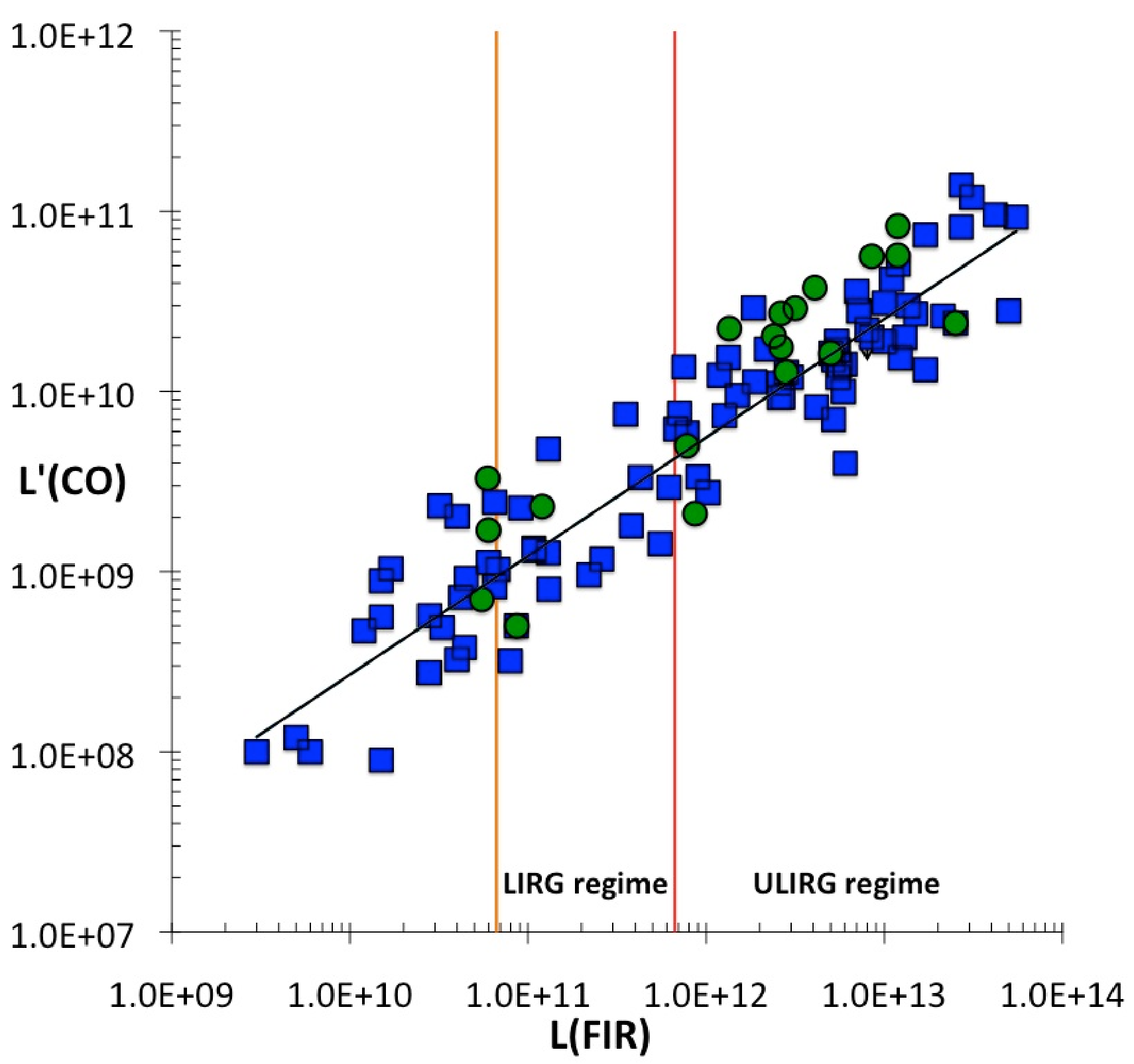}
\includegraphics{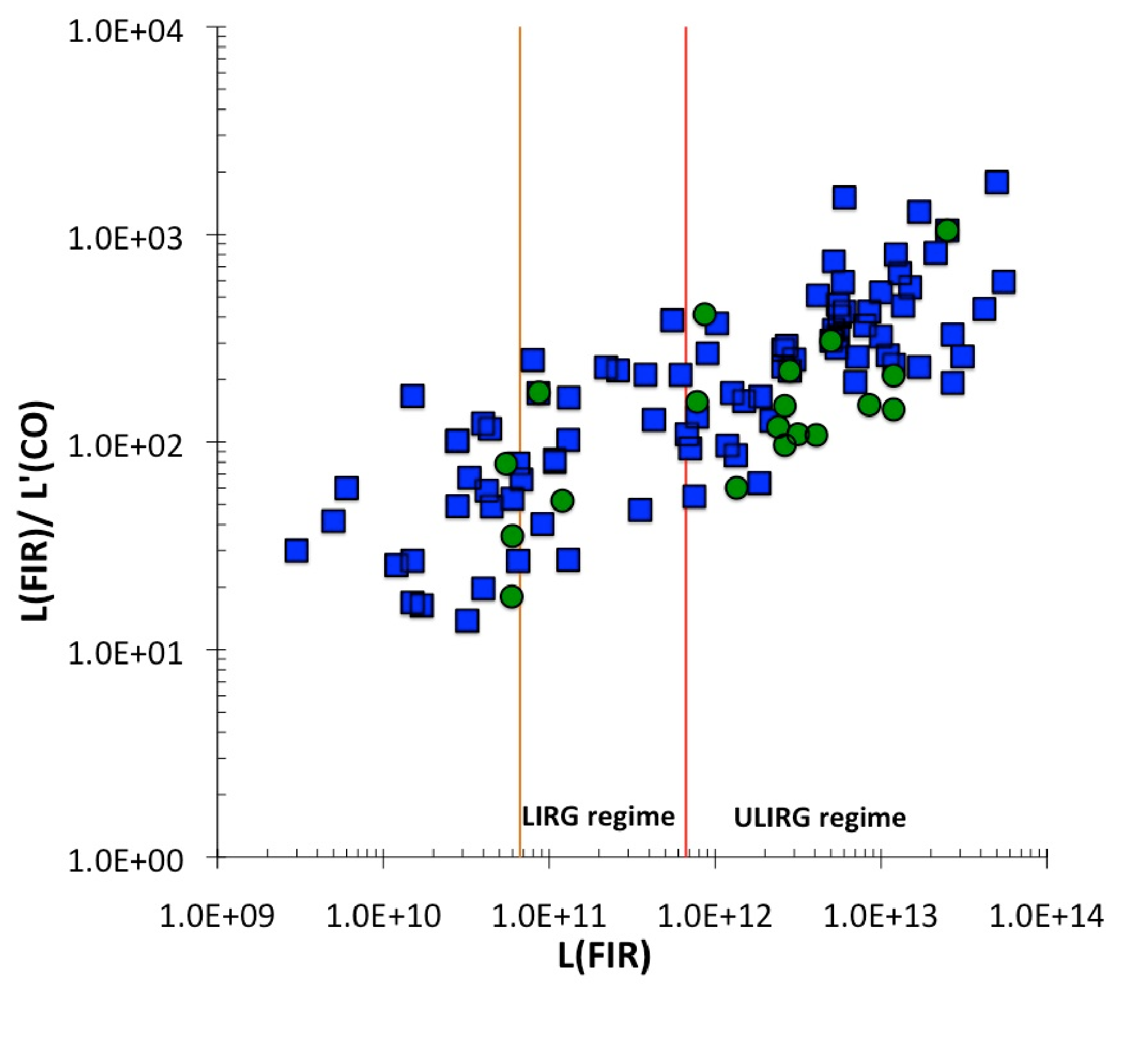}
\includegraphics{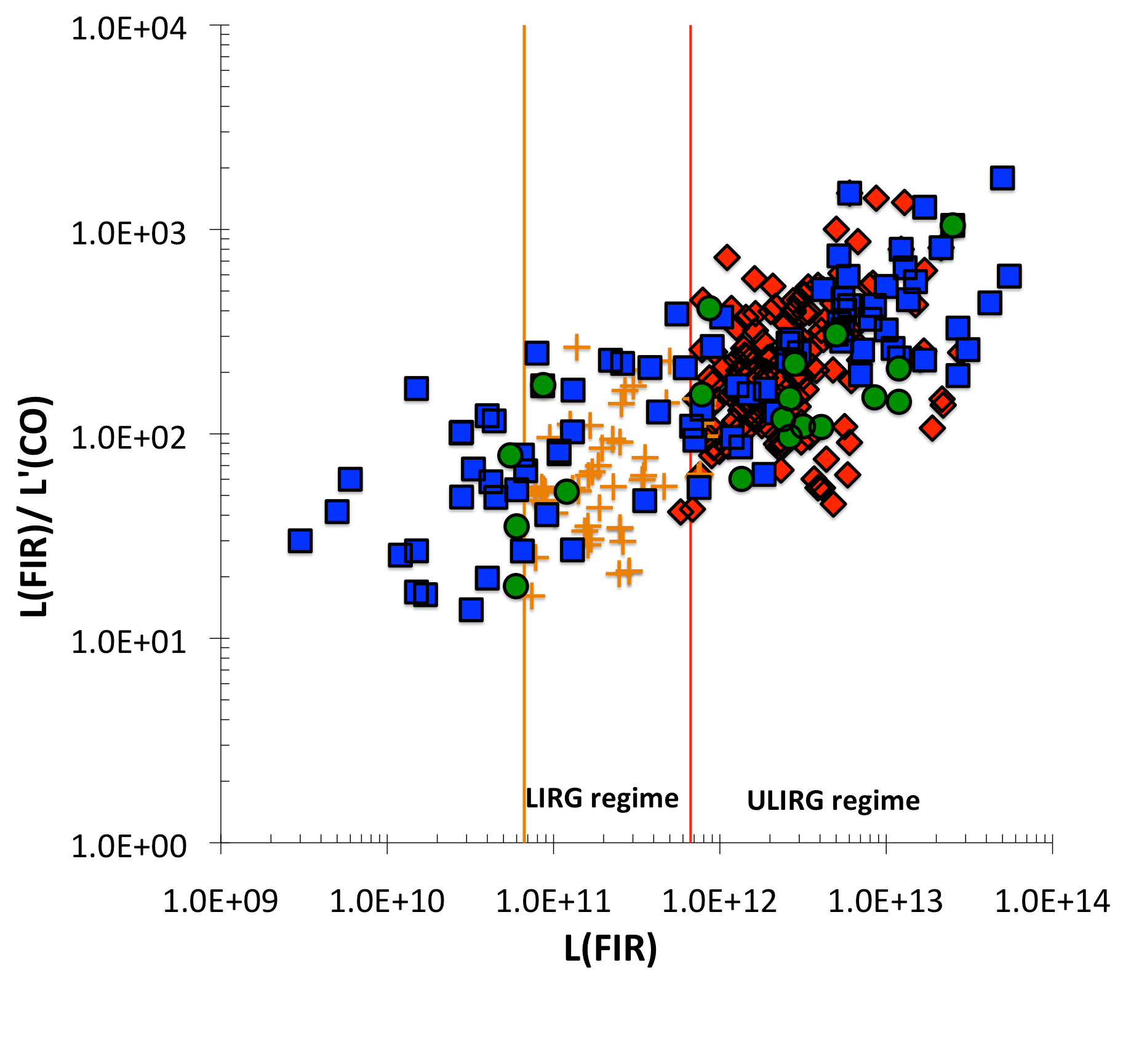}
\vspace{8.4in}
\caption{\lco~vs. \lfir~ (top) for the QSO1 (blue squares) and QSO2 (green circles) in Fig.~3. Upper limits have been excluded.
$\eta = \frac{\lfir}{\lco}$~ vs. \lfir~  only for quasars (middle). LIRGs (orange crosses) and  ULIRGs and high $z$ submm sources (all represented with red diamonds)
are added in the bottom panel. The approximate separation between the different infrared luminosity regimes is indicated with vertical lines.}
\end{figure}

 \subsection{CO kinematics}

CO(1-0) has $FWHM_{ CO}\sim$180-370 km s$^{-1}$ when detected (Table 2) in our QSO2 sample, using the
 $FWHM$ inferred from 1-Gaussian fits to the CO(1-0) line for all objects\footnote{For the double horned profiles of SDSS J1344+05 and 
SDSS J0025-10, 1-Gaussain fits produce FWHM=300$\pm$40 and  265$\pm$20 \kms~ respectively.}. These values are consistent with  KNC12 who measured $FWHM_{ CO}$ in the range $\sim$170-300 \kms.   SDSS J1344+05 and SDSS J0025-10 (Villar-Mart\'\i n et al. \citeyear{vm13}) show evidence of  double horned line profiles, which indicates a diversity of kinematic profiles.  Double peaked  CO lines have been found frequently in different  types of galaxies, including QSO1, radio galaxies, (U)LIRGs  and high z submm galaxies (e.g. Evans et al. \citeyear{ev05}, Bertram et al. \citeyear{ber07}, Oca\~na Flaquer et al. \citeyear{oca10}, Narayanan et al. \citeyear{nar06}, Daddi et al. \citeyear{dad10}).  Rotation is most frequently claimed to explain them,
although  an alternative explanation is mergers.  This is the case of SDSS J0025-10 (Villar-Mart\'\i n et al. \citeyear{vm13}). In this double nuclei merging system, one of the two  CO kinematic component is associated   with the QSO nucleus and/or the intermediate region between the companion nuclei. The other CO component is associated with the northern tidal tail, including a   tidal dwarf galaxy on its tip.

 In general, we find that the CO line is narrower than [OIII]$\lambda$5007, with $\frac{FWHM_{[OIII]}}{FWHM_{CO}}\sim$1-2  (see Table 2).
 This effect is also observed in low $z$ QSO1 (e.g. Shields et al. \citeyear{shi06}).  The  difference in $FWHM$   probably reveals different spatial sizes and geometry  of the ionized and molecular phases and  a higher sensitivity of the [OIII] emission to non gravitational motions, such as outflows. Indeed  \cite{green05} find that  $FWHM_{[OIII]}$ is in general broader   than that of the stars $FWHM_*$ in  a large sample of type 2 AGNs within a broad [OIII]  luminosity range ($\sim$ 3 orders of magnitud, including objects with log(L$_{[OIII]})>$8.3  typical of quasars, see \S2). 
 
   We show in Table 5 $FWHM_{CO}$ (median values) for all the samples with available data at $z\la$0.5\footnote{In those cases where only the $FWZI$ is provided (Bertram et al. \citeyear{ber07}, Scoville et al. \citeyear{sco03}), the ratio $FWZI/FWHM$=1.9 has been assumed. For comparison,  the QSO1 in \cite{xia12} sample  have ratios in the range $\sim$1.4-2.5 with a median value 1.9. }, together with the median values
  of $z$ and  \lfir~ (notice that these can vary relative to Table 4, since here we consider sub-samples with FWHM data and $z\la$0.5). All samples show relatively similar $FWHM_{ CO}^{\rm med}$, in spite of the difference in \lfir. On the other hand,   a trend is hinted for larger $FWHM$ at the highest \lfir~and smaller $FWHM$ at the lowest \lfir.  
  
  $FWHM_{CO}^{\rm med}$ is plotted vs. \lfir~in Fig.~5 for the individual sources.  The same symbols as in Fig.~3 are used, except that now all QSO2 (KNC12 and our sample) are represented with green solid circles. There is no correlation between $FWHM_ {CO}$ and \lfir, but similar trends as in Table 5 are hinted.     The broadest lines ($FWHM_{ CO}^{\rm med}>$400 \kms) are in general only  in ULIRGs, i.e. in the high \lfir~ regime. On the other hand, those objects with the lowest \lfir~  ($\la$several$\times$10$^{10}$\lsun) trend to be associated with narrower CO lines. QSO2 show a range of line widths in the same range as LIRGs (they also have similar \lfir).  The difference might point to the larger incidence of galaxy mergers/interactions at the highest \lfir~, as suggested by the fact that all ULIRGs show signatures of strong interactions and mergers (Sanders \& Mirabel \citeyear{san96})

If we focus on the $z\sim$0.1-0.4 range spanned by the QSO2 sample. The range and median value of $FWHM_{ CO}$ of all 10 QSO2 with available data is similar to that of QSO1 at similar $z$.  In the unification scheme of QSO1 and QSO2, this result suggests that the  CO emitting gas is not coplanar  with the obscuring torus. If this was the case,  the $FWHM$ would depend on the inclination and thus the type 1 vs. type 2 orientation  (see also KNC12). On the other hand, it is not clear what role selection effects are playing, since  CO signals with narrower FWHM are generally easier to detect for a given \lco~ and velocity resolution.      

   $FWHM_{ CO}$ has been often used as a  tracer of dynamical masses in different systems, including quasars (e.g. Bothewell et al. \citeyear{bot13}).  However,  the finding of large reservoirs of molecular gas shifted spatially from the quasar nucleus and associated with companion objects or tidal features shows that this is not always valid (e.g. Villar-Mart\'\i n et al. \citeyear{vm13}, Aravena et al. \citeyear{ara08}, Papadopoulos et al. \citeyear{pap08}; see also Bothwell et al. \citeyear{bot13} for a discussion). It is first  essential to characterize accurately both the kinematics and spatial distribution (size, geometry) of the molecular gas. 

 \begin{table}  
\label{tab:log}      % is used to refer this table in the text
\centering                          % used for centering table
\begin{tabular}{lllllllll}        % centered columns (4 columns)
\hline                % inserts double horizontal lines
 Object   &  $z^{\rm med}$ & Nr. & $FWHM_{ CO}^{\rm med}$    & \lfir$^{\rm med}$  \\      
 class  &   & & \kms   & $\times$10$^{11} \lsun$ \\     \hline
 QSO2 & 0.30 & 10 & 280 &  2    \\
non ULIRG QSO1 &  0.06 & 34 & 223  &  0.6 \\  
ULIRG QSO1 Xia12 &  0.15 & 17 &  275  & 11.9   \\   \hline
LIRGs  & 0.02 &  18 & 253 & 1.4 \\ 
ULIRG  &  0.14 & 91 & 304 & 16.0 \\ 
\hline
\end{tabular}
\caption{Comparison of CO(1-0) $FWHM$  for QSO2 (this work and KNC12), ultraluminous QSO1 (Xia et al \citeyear{xia12}) and
other QSO1 from different samples. Notice that the \lfir~values might vary relative to Table 4 because here we only consider objects with CO(1-0) FWHM measurements. The QSO samples are separated by a horizontal line from (U)LIRGs.}
\end{table}

 \begin{figure}
\includegraphics{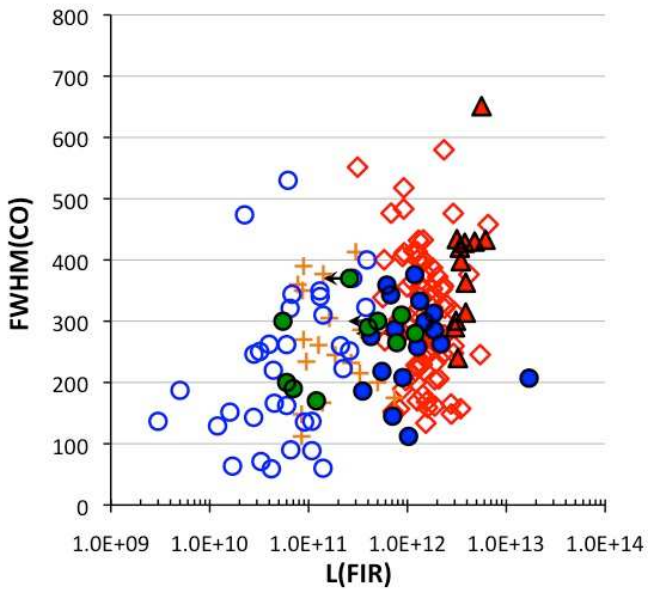}
\vspace{3in}
\caption{$FWHM_ {CO(1-0)}$ in \kms~ vs. \lfir~ for all samples at $z\la$0.5. Symbols as in Fig.~3 except that all QSO2 (KNC12 and this work) are now represented with green solid circles.}
\end{figure}

\section{Summary and conclusions}

We present  results of CO(1-0) spectroscopic observations of 10 SDSS type 2 quasars at $z\sim$0.2-0.3 observed with the 30m IRAM radiotelescope and the  
Australia Telescope Compact Array.  With our work, the total number QSO2   with CO observations at $z<$1.5 increases to $\sim$20, all of which
are at 0.1$\la z \la$ 0.4.

We report 5 new confirmed CO(1-0) detections and 1 tentative detection for our QSO2 sample.  They have \lco$\sim$ several$\times$10$^9$ K km s$^{-1}$ pc$^2$, while upper limits
   for the  non detections are  \lco $<$ 3 $\sigma$ = several$\times$10$^9$ K km s$^{-1}$ pc$^2$.  Assuming a conversion factor $\alpha$=0.8, and including the sample studied by KNC12, the implied
   molecular gas masses for the 20 QSO2 with CO observations at  $z\sim$0.1-0.4 are in the range    \mh2 $\lesssim$4$\times$10$^8$ to $\sim$5$\times$10$^9$ \msun. 

We have constrained the \lir~and \lfir~ of  our sample by fitting the mid to far infrared spectral energy distributions. The \lir~and \lfir~ values of Krips, Neri \& Cox (2012, KNC12)  sample have also been constrained more accurately. Most QSO2 (17/20) are in the LIRG regime or below with \lir$<$10$^{12}$ \lsun. The remaining three have \lir$\sim$10$^{12}$ \lsun, in the transition between the LIRG and ULIRG regimes.  A more complete characterization of the molecular
gas content of QSO2 at similar $z$ requires to expand this study to the highest  \lir~$\ga$several$\times$10$^{12}$ \lsun.
Larger molecular gas reservoirs  \mh2$>$10$^{10}$ \msun~ will   most probably be found. 

We have been able to constrain the \lir/\lfir~ ratios for 9 QSO2. In all cases, this value is in the range $\sim$1.4-1.7 with a median value of 1.5, which is lower than ratios typical of QSO1. This is consistent with a higher obscuration of the mid-infrared luminosity in QSO2 compared to QSO1 as expected.

   At intermediate $z$ no difference is found on \lco~ (or the molecular gas content) between QSO2 and QSO1 once the infrared luminosities are accounted for. This is consistent with the unification model of QSO1 and QSO2.

   QSO2 fall on the \lco~ vs. $z$,  \lco~ vs. \lfir~ and  $\eta=\frac{L_{FIR}}{L_{CO}}$ vs. \lfir~correlations defined by  quasars at different $z$.  The location of
   the QSO2 in these diagrams is discussed  in comparison with samples of  QSO1, LIRGs, ULIRGs and high $z$ submm sources.

 CO(1-0) has $FWHM_{ CO}\sim$180-370 km s$^{-1}$ when detected, with a variety of kinematic profiles (single or double horned). 
In our sample, the CO line is in general narrower than the [OIII]$\lambda$5007, as observed in low $z$ QSO1.  This probably reveals different spatial sizes and geometry of the ionized and molecular phases and a higher sensitivity of the [OIII] emission to non gravitational motions, such as outflows.  The range and median value of $FWHM_{ CO}$ of all 10 QSO2 with available data is similar to that of QSO1 at similar $z$, although this result is tentative. In the unification scenario of QSO1 and  QSO2 this result, if confirmed, suggests that  that the spatial distribution of the CO(1-0) emitting gas is not related to the obscuring torus and is therefore independent of its orientation relative to the observer.

     To perform a more complete and adequate comparison between QSO1 and QSO2 it is essential to expand this study in $z$ and \lfir, for both QSO1 and very specially QSO2. The low redshift ($z<$0.1) range  is completely unexplored for low $z$ QSO2, as well as the ULIRG regime at intermediate z. Similarly, it will be useful to enlarge the sample of non-ULIRG QSO1 at $z >$0.2 to ensure an overlap on both $z$ and \lfir~ with the QSO2 samples.

\section*{Acknowledgments}
This work has been funded with support from the Spanish former
Ministerio de Ciencia e Innovaci\'on through the grant AYA2010-15081. MR acknowledges support by the Spanish MINECO through grant AYA
2012-38491-C02-02, cofunded with FEDER funds. Thanks to the staff at IRAM Pico Veleta for their support during the observations.  IRAM is supported by INSU/CNRS (France), MPG (Germany) and
IGN (Spain). The Australia Telescope is funded by the Commonwealth of Australia for operation as a National Facility managed by CSIRO.  
 This research has made use of the VizieR catalogue access tool (CDS, Strasbourg, France)  and the NASA/IPAC Extragalactic Database (NED). 

The SDSS spectra and images of the sample have been used for different  issues essential to the paper.    Funding for the SDSS and SDSS-II has been provided by the Alfred P. Sloan Foundation, the Participating Institutions, the National Science Foundation, the U.S. Department of Energy, the National Aeronautics and Space Administration, the Japanese Monbukagakusho, the Max Planck Society, and the Higher Education Funding Council for England. The SDSS Web Site is http://www.sdss.org/.    The SDSS is managed by the Astrophysical Research Consortium for the Participating Institutions.

\appendix

\section[]{Fits of the Spectral Energy Distributions}

The figures with the mid (WISE; 3.3, 4.6, 11.6, 22.1 $\mu$m) to far infrared  (IRAS; 60 and/or 100$\mu$m)  photometric data and the SED fits for our sample are shown here.
The solid diamonds correspond to the detected fluxes,  while open triangles indicate upper limits.  When  two values appear at  $\lambda\sim$22-25$\mu$m, these correspond to the WISE 22.1$\mu$m  and IRAS 25$\mu$m bands. See \S4.3 for more detailed information. 

\begin{figure*}
\includegraphics{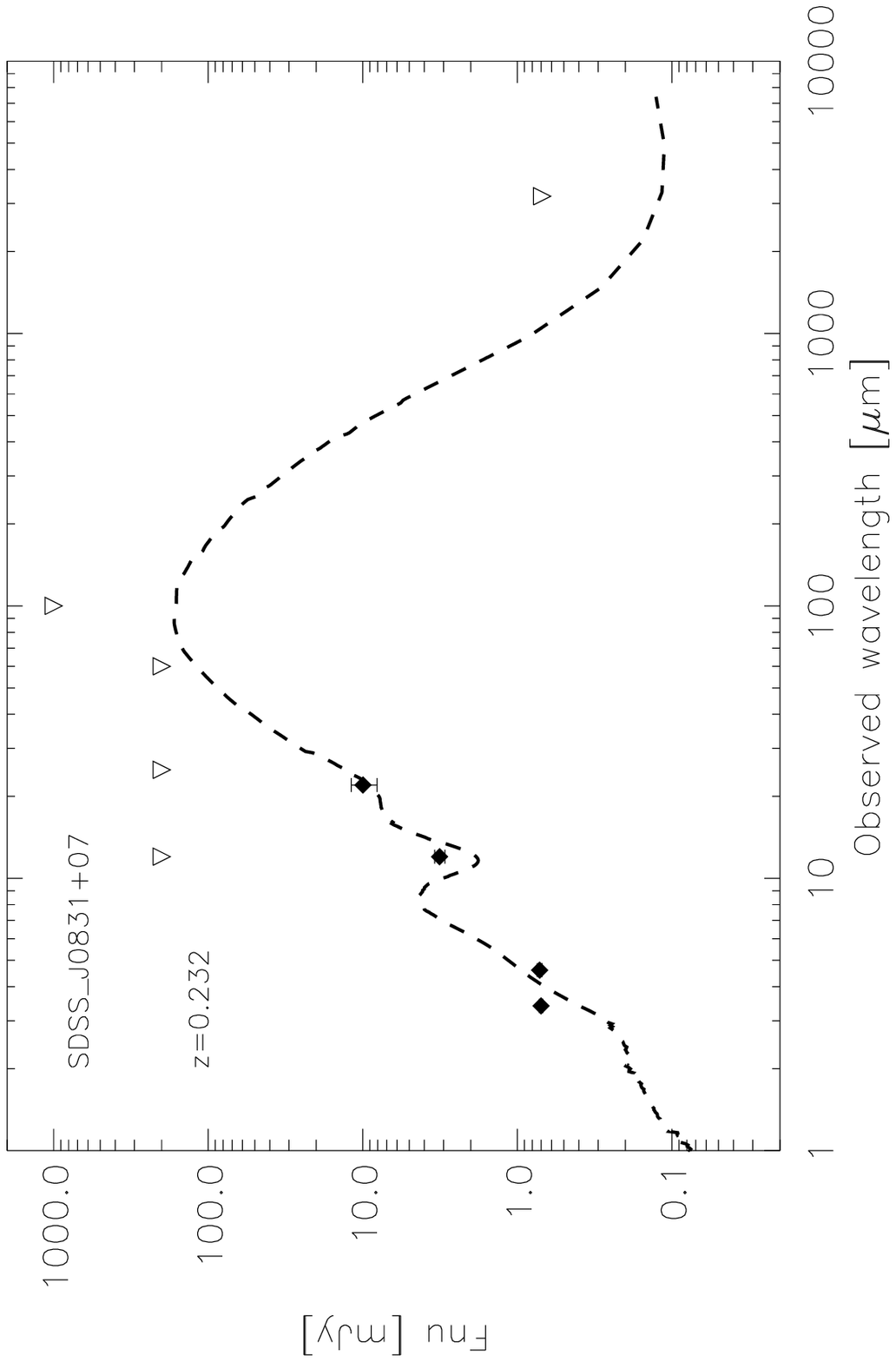}
\includegraphics{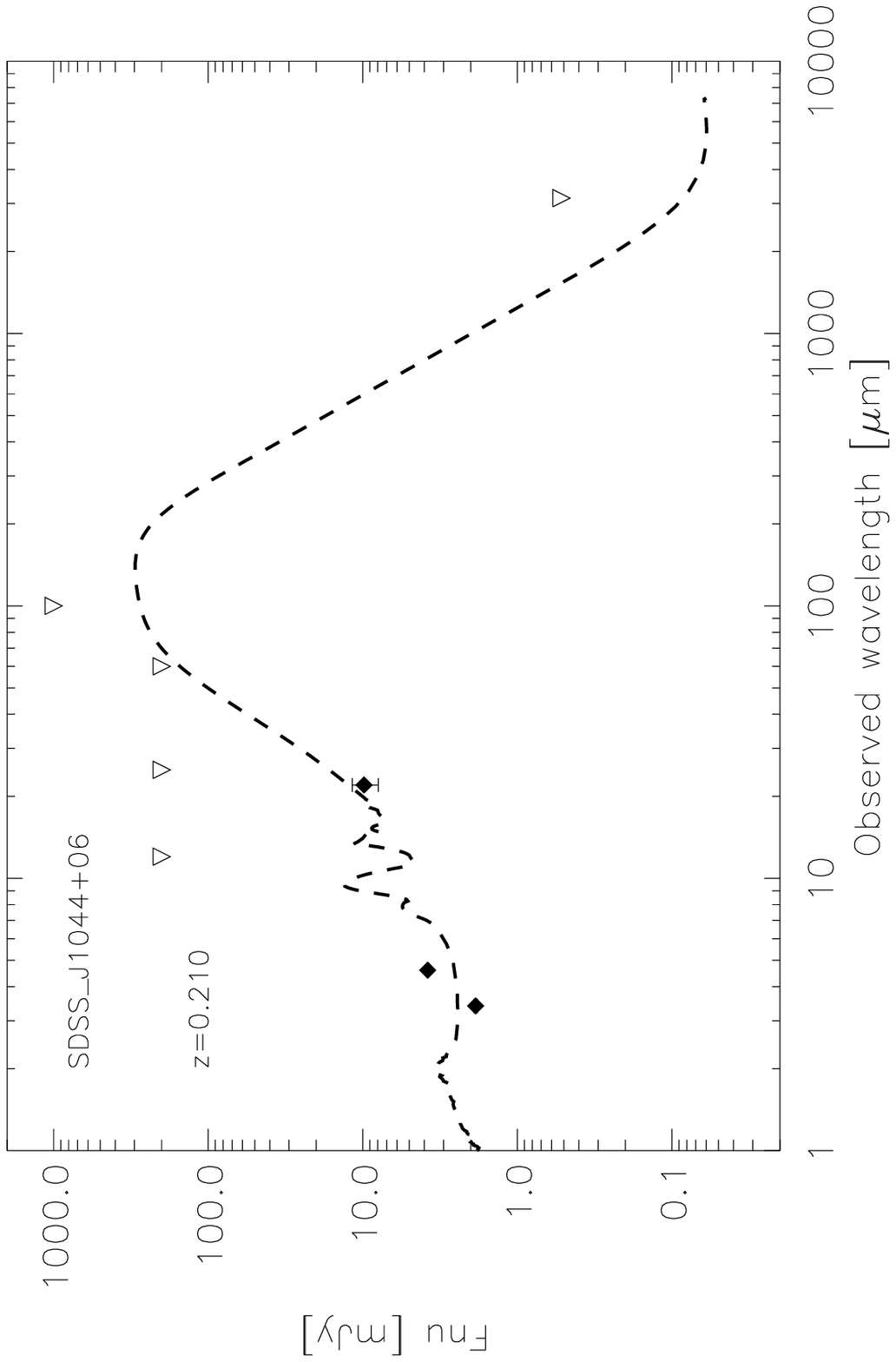}
\vspace{2.5in}
\includegraphics{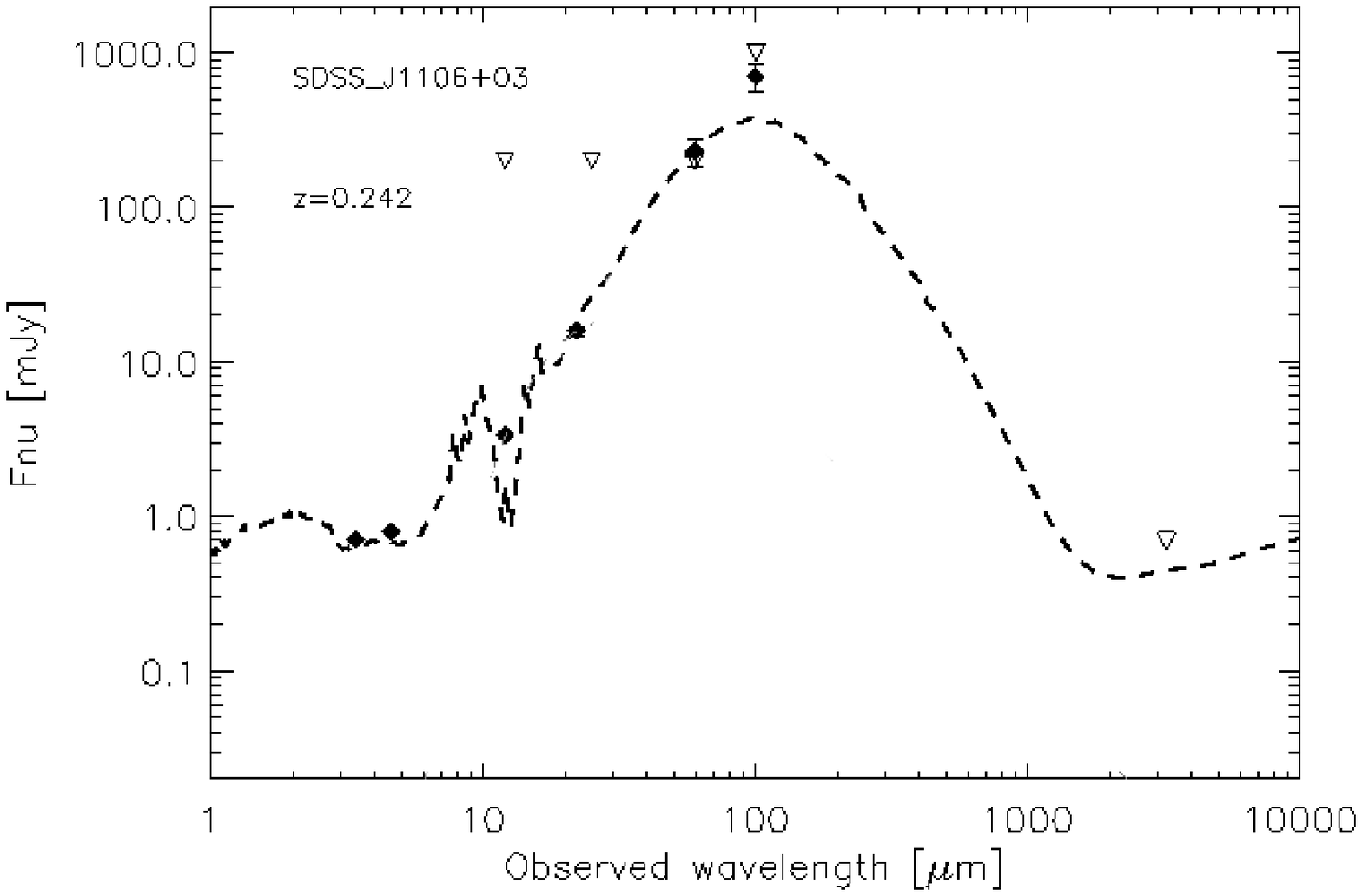}
\includegraphics{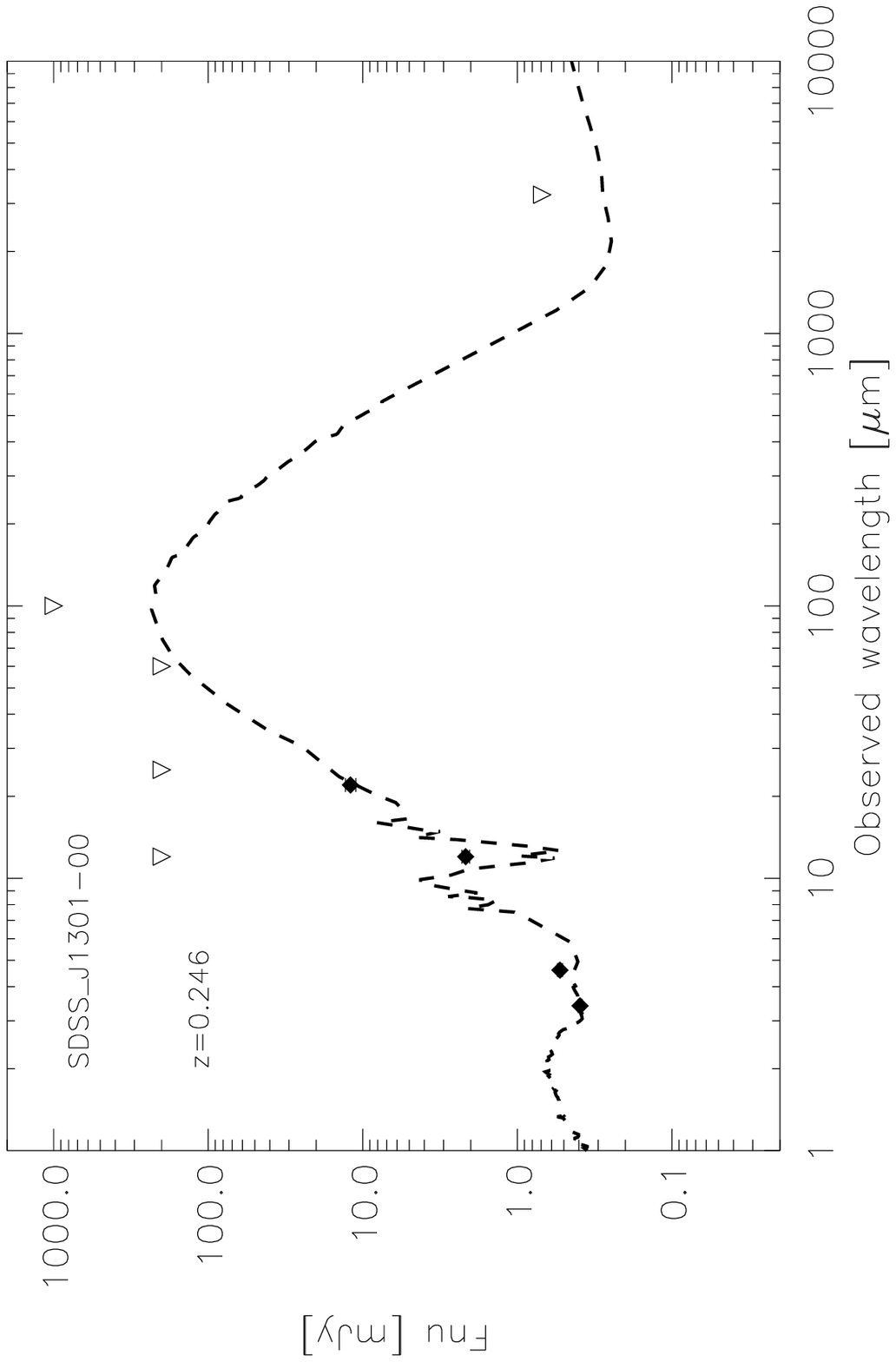}
\vspace{2.5in}
\includegraphics{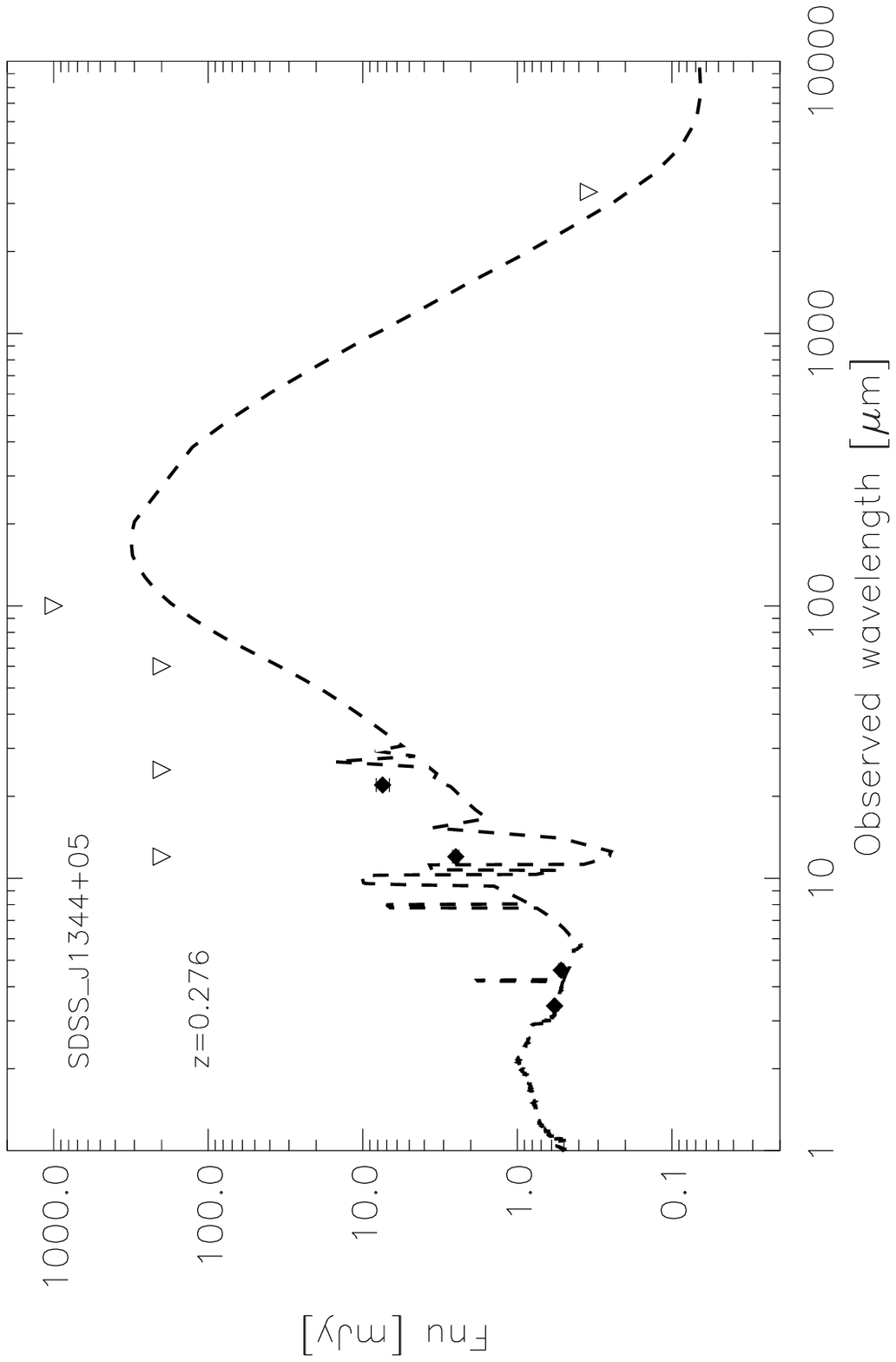}
\includegraphics{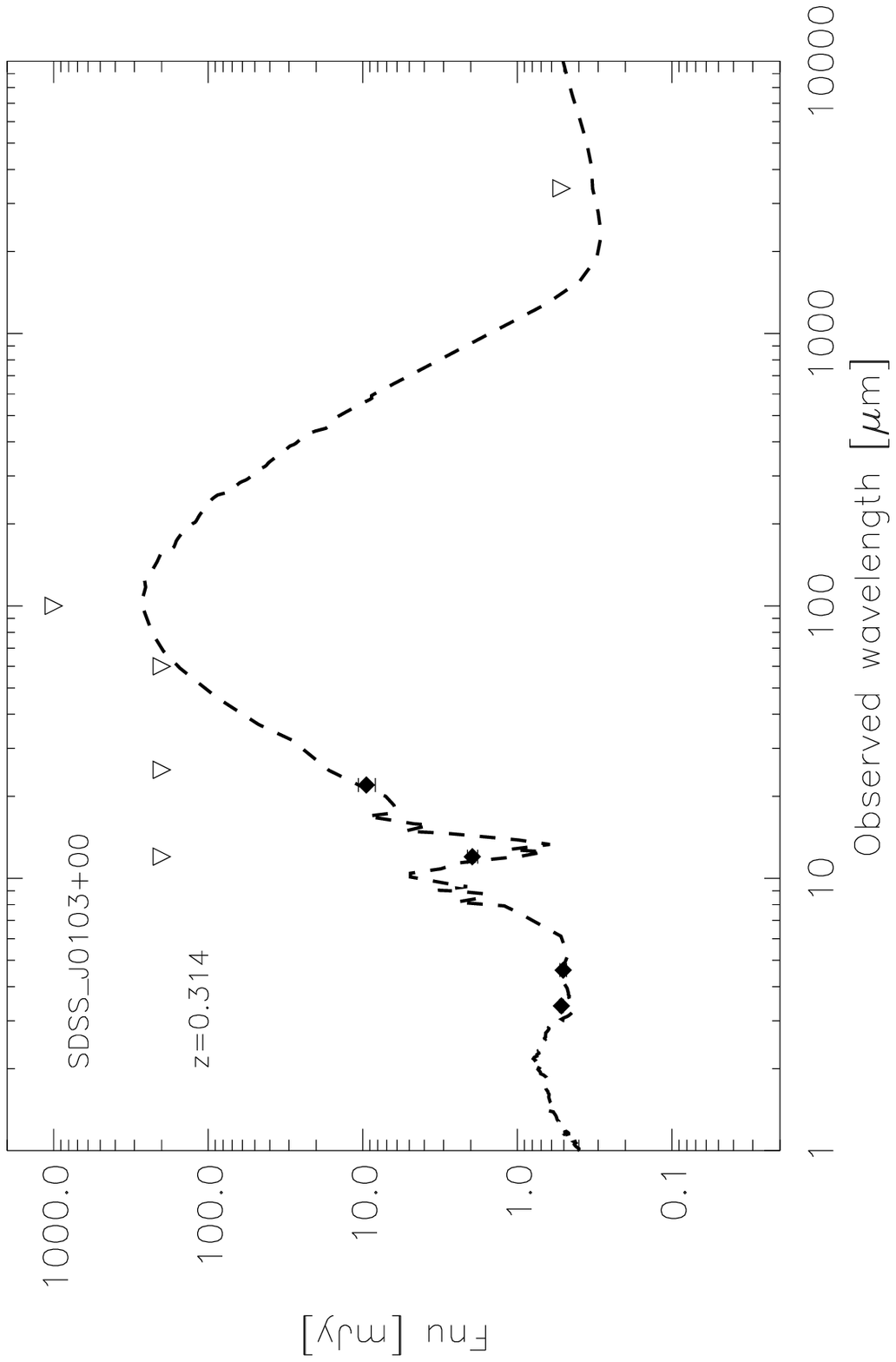}
\vspace{2.8in}
\caption{Fits of the infrared SEDs of the QSO2 in our sample. Detections and upper limits are shown with solid diamonds
and open triangles respectively. For those objects with no detections at far infrared walengths (60 and/or 100 $\mu$m) only upper limits for \lir~and \lfir~ 
could be stablished. For these, the fits producing the maximum possible \lir~  consistent with the mid-infrared photometry and the far
infrared upper limits are shown.} 
\end{figure*}

\newpage
~~

\begin{figure}
\includegraphics{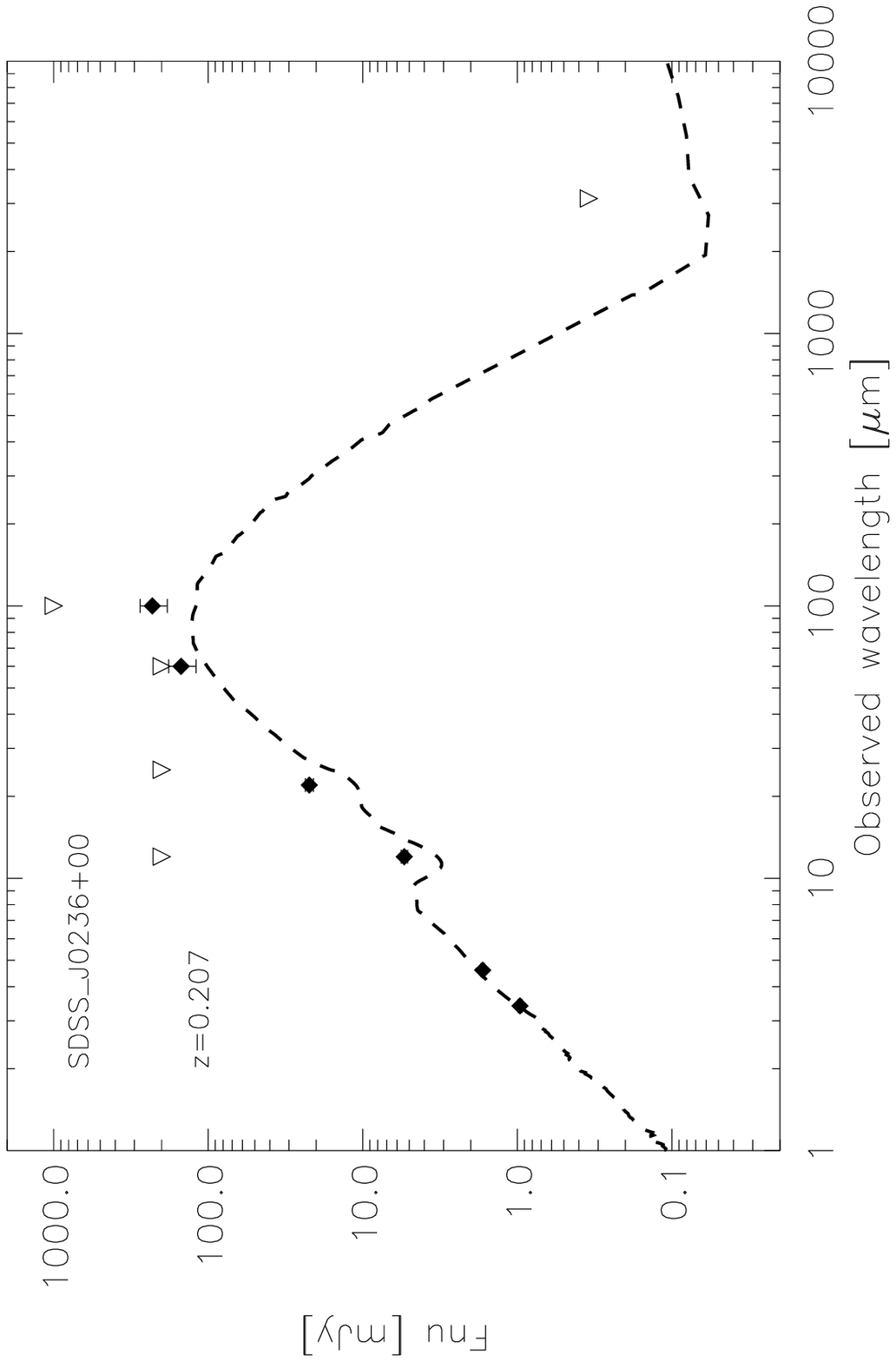}
\includegraphics{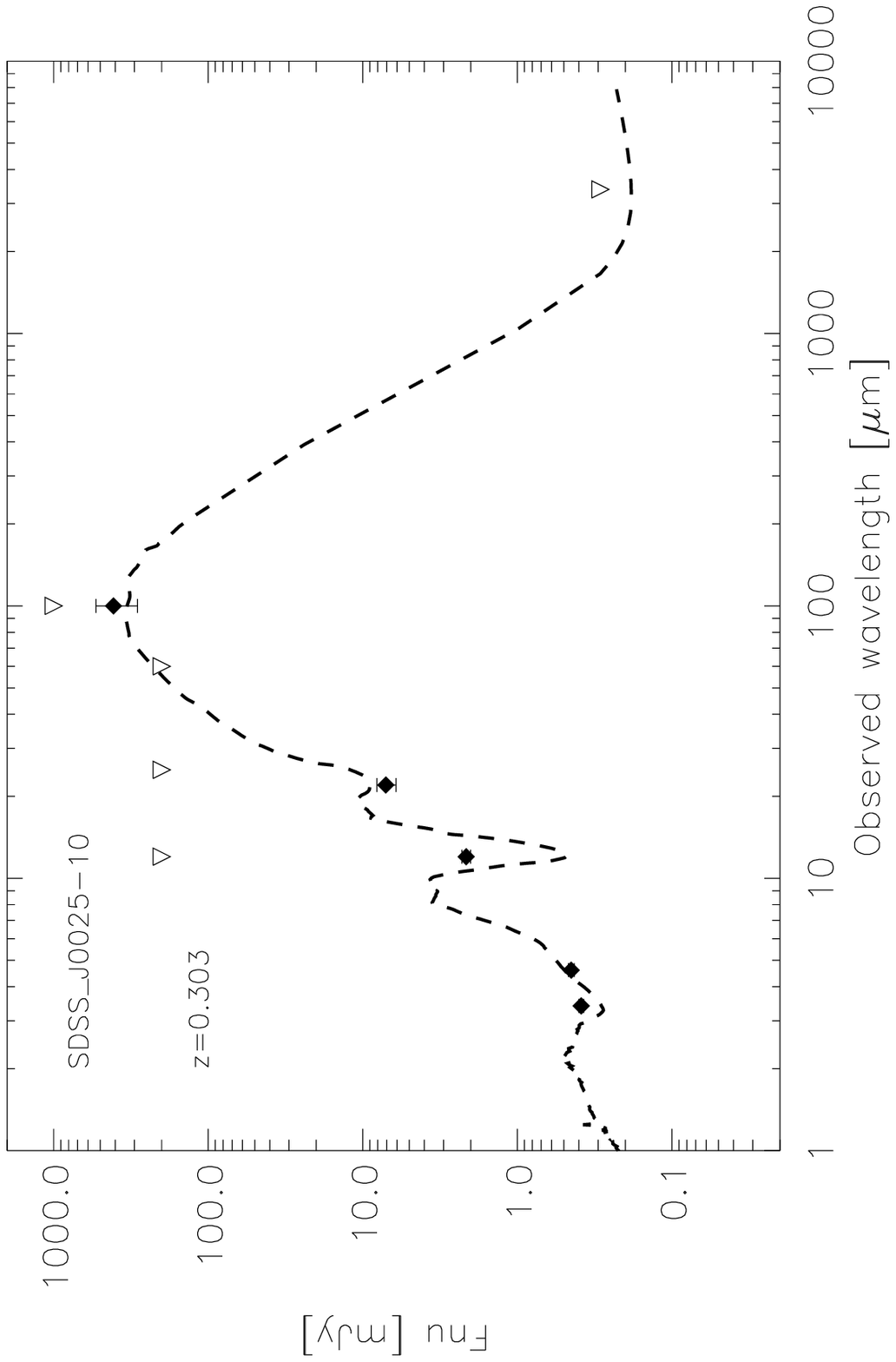}
\includegraphics{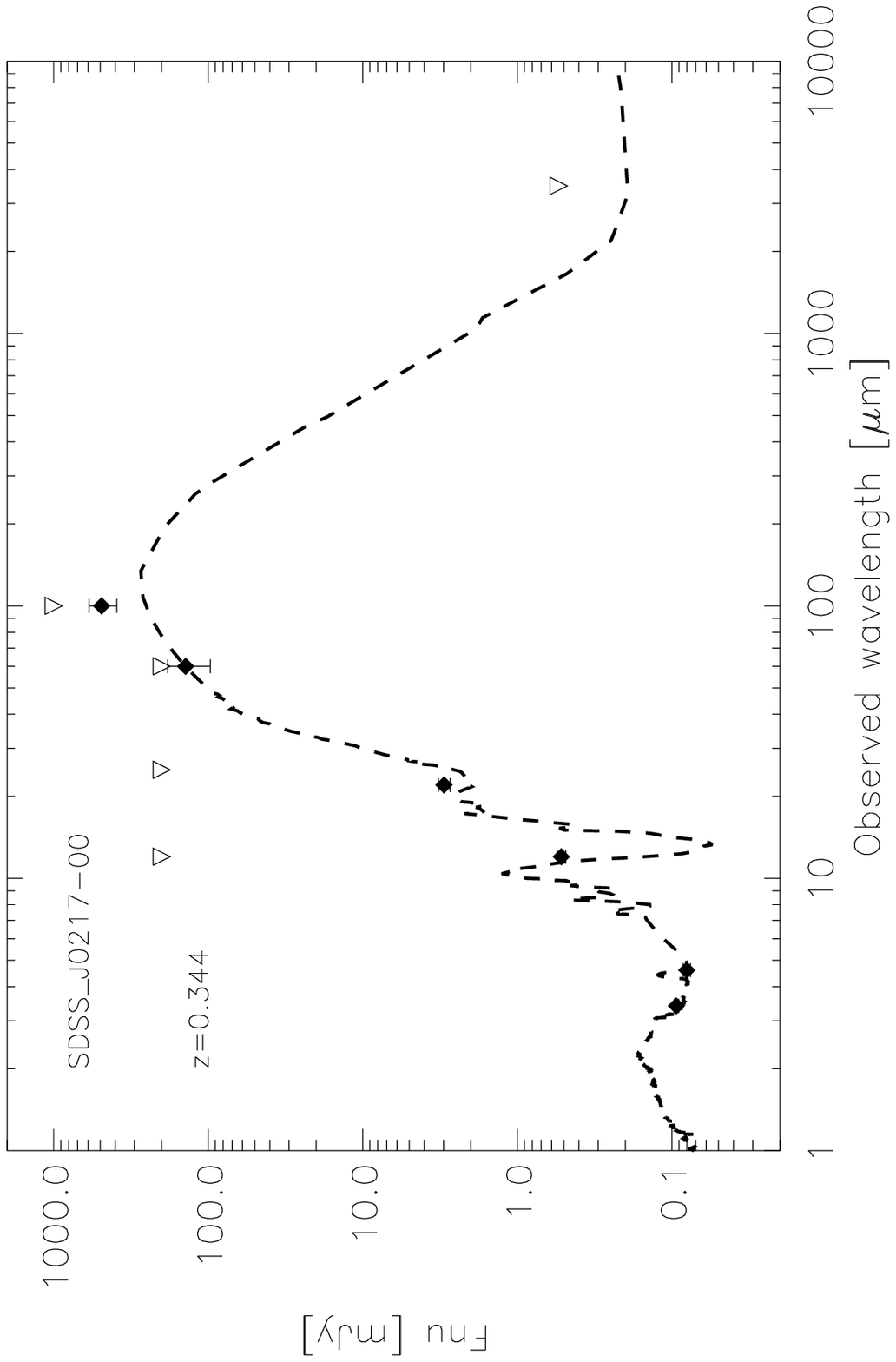}
\vspace{5.5in}
\caption{Continuation of Fig. A1.}
\end{figure}

  \end{document}